%% file: ms_astroph2.tex
\documentclass[12pt,preprint]{aastex}

\usepackage{emulateapj5}

\newcommand{\ra}[3]{${\rm RA}={#1}^{\rm h}{#2}^{\rm m}{#3}^{\rm s}$}
\newcommand{\dec}[3]{${\rm Dec.}={#1}^{\circ}{#2}\arcmin{#3}\farcs$}

\newcommand{\e}[1]{$10^{#1}$}
\newcommand{\ee}[1]{$\times 10^{#1}$}
\newcommand{\cm}[1]{~cm$^{#1}$}
\newcommand{\cms}{~cm$^{-3}$\,s}
\newcommand{\kms}{~km\,s$^{-1}$}

\newcommand{\msol}{$M_{\odot}$}

\shorttitle{HETGS observation of Cas A}
\shortauthors{Lazendic et al.}

\begin{document}

\title{The Kinematic and Plasma Properties of X-ray Knots in Cassiopeia~A
    from the Chandra HETGS  }

\author{J. S. Lazendic, D. Dewey, N. S. Schulz, and C. R. Canizares}
\affil{MIT Kavli Institute for Astrophysics and Space Research, Cambridge MA 02139}

\begin{abstract}

We present high-resolution X-ray spectra from the young supernova
remnant Cas A using a 70-ks observation taken by the {\em
Chandra} High Energy Transmission Grating Spectrometer
(HETGS).  Line emission, dominated by Si and S ions, is used for
high-resolution spectral analysis of many bright, narrow regions of
Cas A to examine their kinematics and plasma state. 
These data allow a 3D reconstruction using the 
unprecedented X-ray kinematic results: we derive
unambiguous Doppler shifts for these selected regions, with values
ranging between $-2500$ and $+4000$\kms\ and the typical velocity error less than 200\kms.   Plasma diagnostics of these
regions, derived from line ratios of resolved He-like triplet lines
and H-like lines of Si, indicate temperatures largely around 1
keV, which we model as O-rich reverse-shocked ejecta.  The ionization
age also does not vary considerably over these regions of the remnant. The gratings
analysis was complemented by the non-dispersed spectra from the same
dataset, which provided information on emission measure and elemental
abundances for the selected Cas~A regions. The derived electron density of
X-ray emitting ejecta varies from 20 to 200\cm{-3}. The measured abundances of
Mg, Si, S and Ca are consistent with O being the dominant element in the Cas~A plasma.  
With a diameter of 5\arcmin, Cas~A is the largest source
observed with the HETGS to date. We, therefore, describe the 
technique we use and some of the challenges we face in the HETGS data
reduction from such an extended, complex object.

\end{abstract}


\section{Introduction}

Cassiopeia~A (Cas~A, G111.7--2.1) is the youngest Galactic supernova
remnant (SNR), believed to be a product of a SN explosion in $\sim
1670$ \citep[e.g.][]{thorstensen01,fesen06}.  The SNR is the brightest object
in the radio band, and is very bright in the X-ray band, providing a suitable target for detailed studies across a wide range of wavelengths. Cas~A shows a
wealth of phenomena important for studying the SN progenitor, its
explosion mechanism, the early evolution of SNRs, and their impact
 on the interstellar medium.

The remnant appears as a bright, clumpy ring of emission with a
diameter of around 3\arcmin\, associated with the SNR ejecta, while
the fainter emission from the SNR forward shock forms a filamentary
ring with a diameter of 5\arcmin.  The almost perfectly circular
appearance of the SNR is disrupted by the clear extension in the NE
region of the SNR, called the jet, 
and a less obvious one in the SW called the counter-jet \citep[e.g][]{fesen96,hwang00}.

Many of Cas~A's bright knots have been identified as
shocked ejecta, still clearly visible due to the young age of the SNR. 
Observations in the optical band provided the first insights into the
kinematics and chemical composition of these ejecta 
\citep[e.g.][]{minkowski59,peimbert71,chevalier78}.  The
optical-emitting ejecta have been classified into two main groups: the
fast-moving knots (FMKs), moving with speeds from 4000\kms\ up to
15,000\kms, and the slow-moving quasi-stationary floculi (QFS), moving
with speeds less than 300\kms\ \citep{kamper76}. The emission from
FMKs, believed to be ejecta, is H-deficient (from the lack of H$\alpha$ emission) 
and dominated by forbidden O and S emission \citep[e.g.][]{fesen01}. 
This lack of hydrogen in FMKs suggests that Cas A was a Type Ib supernova produced by the
core-collapse of a Wolf-Rayet star \citep[e.g.][]{woosley93}.
The emission from QFSs is rich in N,
 which is the reason these knots are believed to originate from the circumstellar envelope released by the progenitor star and subsequently shocked (or ``over-run'').

While the available kinematic information from the optical observations 
of Cas~A is based on observations of thousands of individual knots, X-ray observations 
are more dynamically important because X-rays probe a much
larger fraction of the ejecta mass (more than 4\msol) than does 
the optical emission (accounted for less than 0.1\msol).
Kinematics in the X-ray band have been studied using radial velocities \citep{markert83,holt94,hwang01,willingale02} and proper motion \citep{vink98,delaney03,delaney04}, but 
there is still a need for high-resolution spatial and spectral observations that will 
match the quality of results provided in the optical band. 
Here we present observations of Cas~A with the {\em Chandra} High
Energy Transmission Grating Spectrometer (HETGS). Cas~A is one of the
rare extended objects viable for grating observations due to its
bright emission lines, narrow bright filaments and small bright clumps
that standout well against the continuum emission.

\section{Observations and Data Analysis}

Cas~A was observed with the HETGS on board the {\em Chandra X-ray
Observatory} in May 2001 as part of the HETG guaranteed time observation
program (ObsID 1046). The exposure was 70~ks, the roll angle was
86\degr\ West to North and the aim point was at \ra{23}{23}{29},
\dec{+58}{48}{59}{6}.
The HETGS consists of two grating arms with different dispersion directions: 
1) the medium-energy grating (MEG) arm which covers an 
energy range of 0.4-5.0~keV and has FWHM of 0.023 \AA, and 2) the high-energy grating 
(HEG) arm which covers 0.9-10.0 keV band and has FWHM of 0.012 \AA\ (for more details see
e.g. \citealt{canizares05}). The different dispersion directions and
wavelength scales of the two arms provide a way to resolve potential
spectral/spatial confusion problems \citep[see Appendix A; also][]{dewey02}. 
The HETGS is used in conjunction with the Advanced CCD Imaging
Spectrometer (ACIS-S). In addition to the dispersed images, a non-dispersed
zeroth-order image from both gratings 
occupies the S3 chip at the aimpoint, while the
dispersed photons are distributed across the entire S-array. The zeroth-order
image, thus, has spatial and spectral resolution provided by the ACIS
detector.

The Cas~A HETGS event file was re-processed using the standard
procedures in CIAO\footnote{http://asc.harvard.edu/ciao/} 
software version 3.2.2, employing the latest
calibration files. Processing of the data included removal of the
pipeline afterglow correction (a significant fraction of the rejected
events was from the source), correctly assigning ACIS pulse heights to
the events and filtering the data for energy, status, and grade. To
retain more valid events, the removal of artifacts (destreaking) from
the S4 chip was done by requiring more than 5 events in a row in order
to destreak it.

The standard procedures in CIAO for gratings data assume that the
source is point-like. We, therefore, used alternative software
(written in IDL) that basically follows the steps of the CIAO threads for
gratings spectra, but accounts for an extended, filament-like source
during extraction of the PHA spectra and the corresponding spectral 
redistribution matrix files (RMFs). The RMFs are particularly 
important as they relate the photon energy scale to the detector dispersion 
scale of the gratings. We also used standard CIAO threads to create 
auxiliary response files (ARFs), which contain the information on telescopes effective
area and the quantum efficiency as a function of energy averaged over
time.  The resulting ARFs were examined for bad columns, and the parts
of the spectrum where the bad columns are present were ignored in the
fitting procedure. Details of this ``filament analysis'' are presented
in Appendix~A.

For the zeroth-order data, we extracted spectra and corresponding ARFs and CCD RMFs with 
 the standard CIAO thread {\tt acisspec}. For the background spectra we tried using the emission-free region on the S3 chip, the SNR regions surrounding our bright knots, and the regions from the ACIS blank-sky event files. We found no difference when using any of these spectra, so we decided to use the ACIS blank-sky events. The background spectra are extracted with region sizes a factor of 2 larger than the source spectra. The zeroth-order model fits were carried out binning the data to contain a minimum of 25 counts per bin.

\section{Results}

\begin{figure}
\centering
\includegraphics[height=20cm]{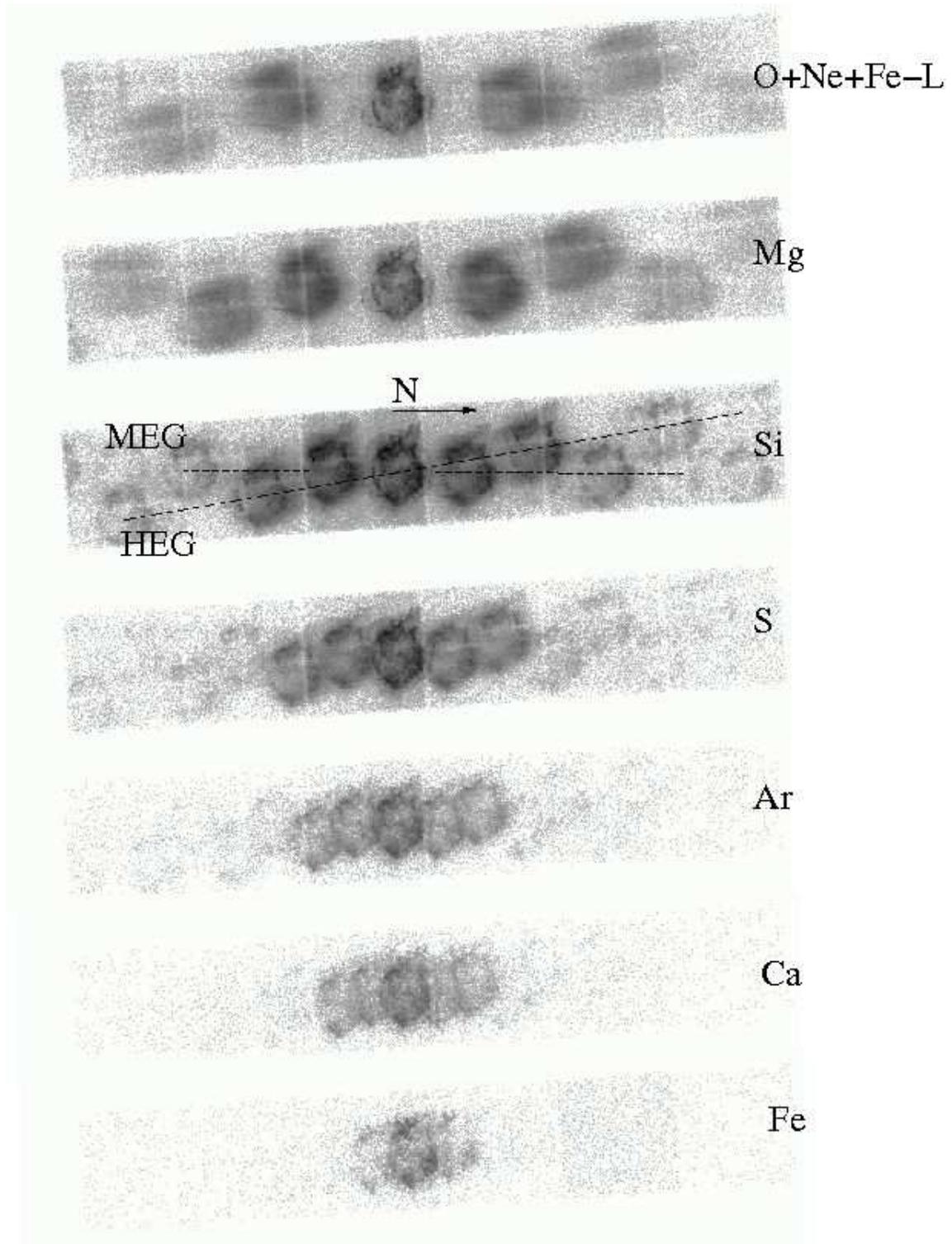}
\caption{HETG data of Cas A: images of the different line bands. 
 Grating dispersion axes and north direction are marked in the Si-band image.
Note the smeared out dispersed-order images, especially
in the O+Ne+Fe-L band, due to multiple lines and velocity shifts
which, however, do not affect the central, zeroth-order image. }
\label{fig-casA}
\end{figure}

Figure~\ref{fig-casA} shows the non-dispersed (zeroth-order) image of Cas~A at the center of each panel and dispersed images in different energy bands that contain predominately He- and H-like ions, and in the case of Fe also Li- and Be-like ions: O+Ne+Fe-L
(0.65--1.2~keV), Mg (1.25--1.55~keV), Si (1.72--2.25~keV), S
(2.28--2.93~keV), Ar (2.96--3.20~keV), Ca (3.75-4.00~keV) and Fe-K
(6.30--6.85~keV).  The Si-band image also indicates the two dispersion axes of the MEG and HEG, which are rotated against each other by $\sim10\degr$. The Cas~A X-ray spectrum is
dominated by Si and S lines and these bands produce the clearest dispersed images. 
The dispersion angle $\theta$ defines the location of the dispersed photon and varies 
 nearly linearly with wavelength $\lambda$, $sin \theta = m\lambda / p$,  where
$m$ is the order of dispersion and $p$ is the period of the grating.  Thus, the dispersed images in the longer wavelengths (O+Ne+Fe-L band) are spread out furthest across the spectroscopy CCD array, while in the shorter wavelength (Fe-K band) the images stay tight to the zeroth-order. Because of this, the longer-wavelength images are more sensitive to velocity gradients and distortions 
are more prominent, as indicated by the larger smearing seen in the top two panels.
 The O+Ne+Fe-L image is
additionally confused due to the many lines in this band \citep[e.g., Fe-L lines
and the O~Lyman series; see][]{bleeker01},  and a larger relative contribution by the continuum.
 
\begin{figure}
\centering
\includegraphics[height=12cm]{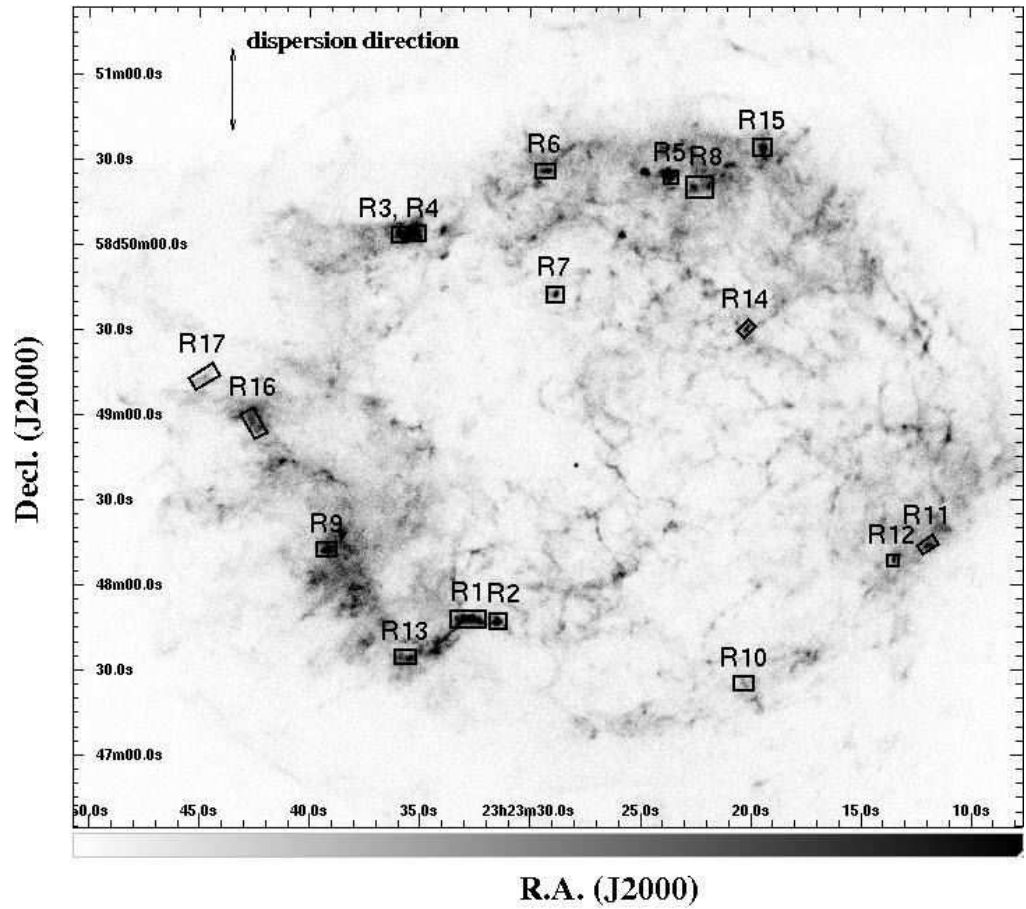}
\caption{ Zeroth-order Cas~A image from this work. Regions used in the analysis of HETGS Cas A data. These regions are spatially
narrow and they are isolated sufficiently above the local and extended background to provide a clear line profile for spectral fitting.}
\label{fig-regions}
\end{figure}

\begin{figure}
\centering
\includegraphics[height=25cm]{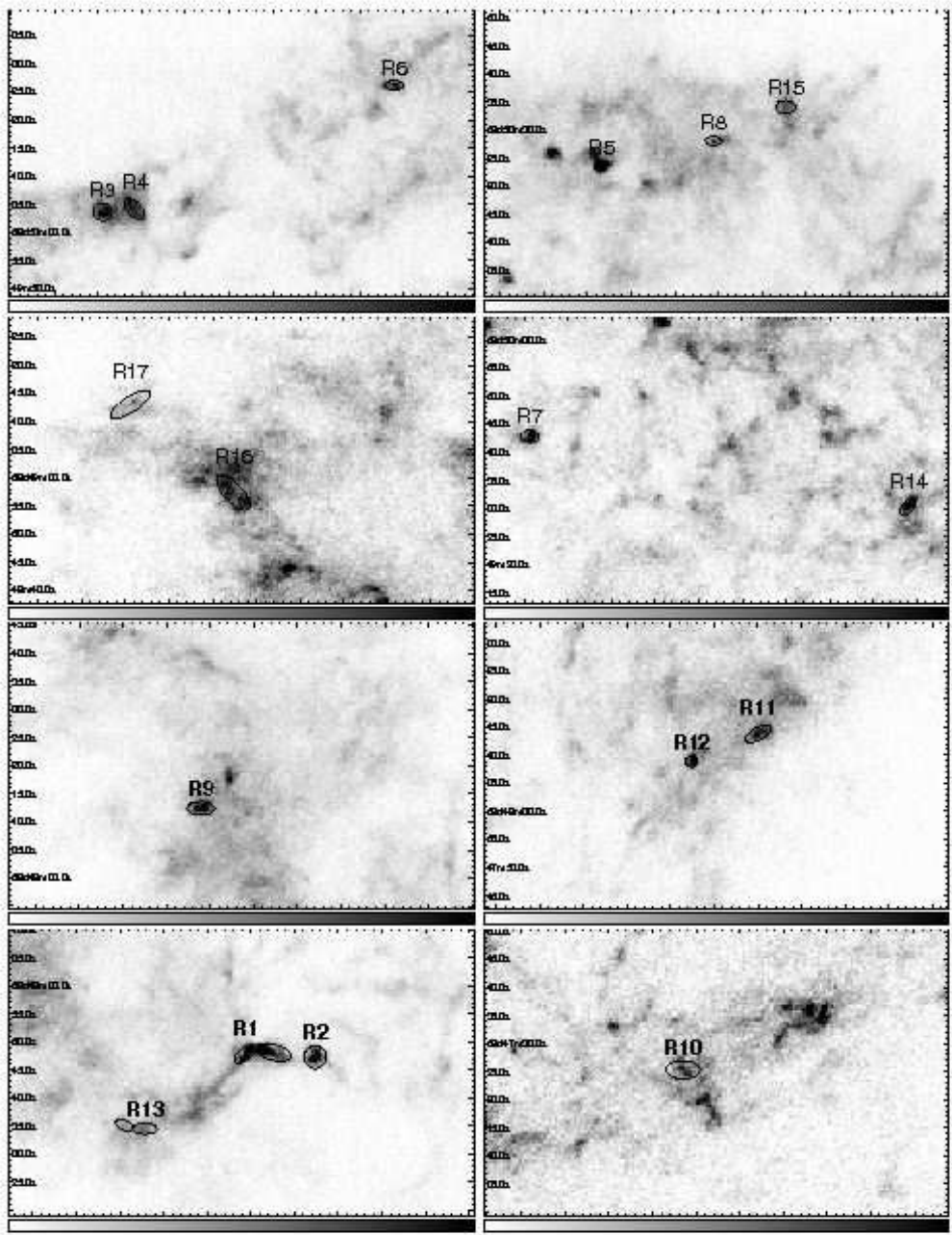}
\caption{Close up of the regions used in the analysis of HETGS Cas A data. Note that the grayscale dynamic range is not the same for all the sub-plots. Some of the regions have been covered with two elliptical regions to follow the HETGS extraction path (see Appendix~A).}
\label{fig-close-up}
\end{figure}

\begin{figure}
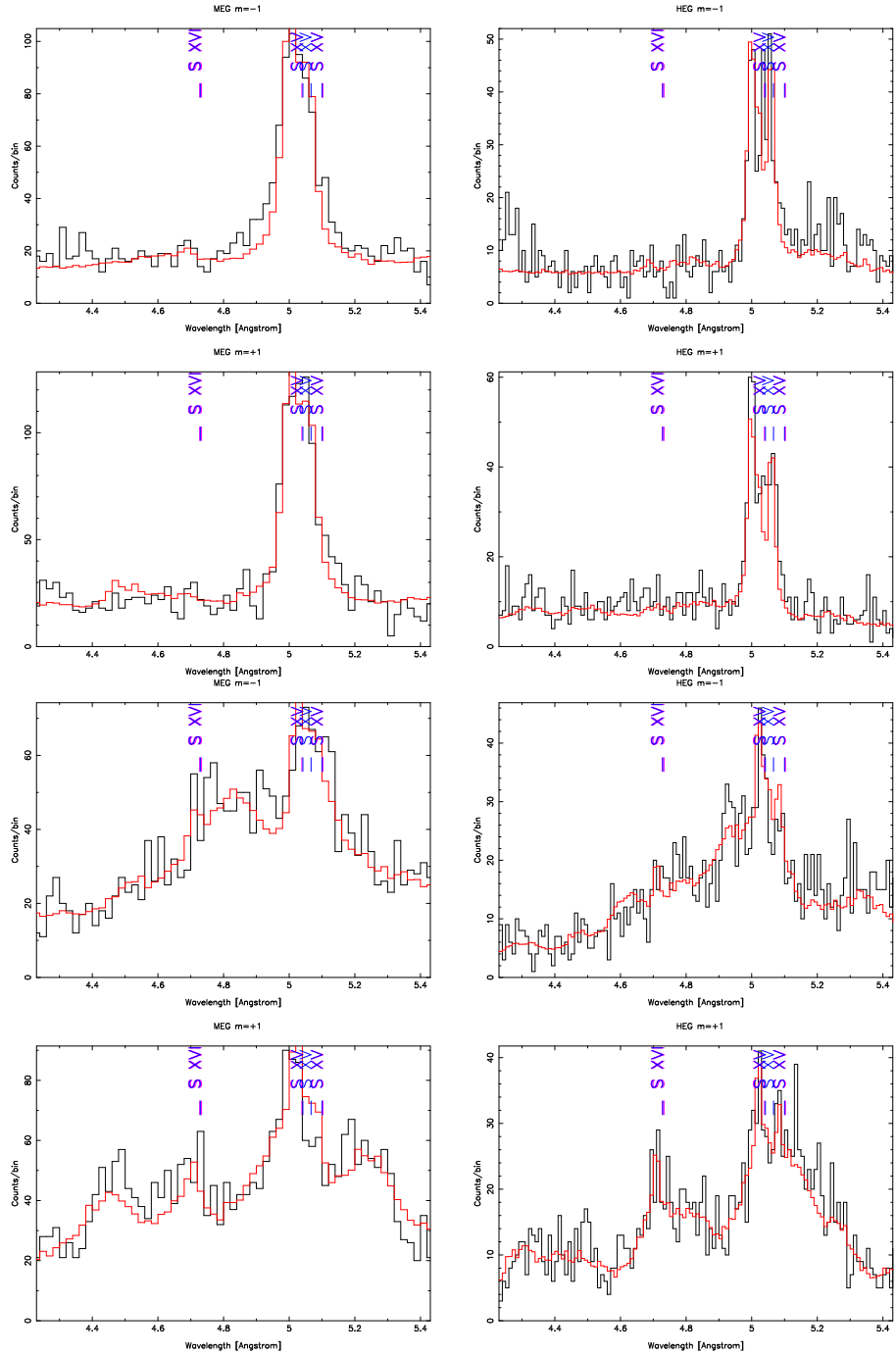

\centering
\includegraphics[height=12cm,angle=-90]{f5a.ps}
\includegraphics[height=12cm,angle=-90]{f5b.ps}\\
\caption{Examples of HETGS spectra of S band encompassing four S lines used in Cas~A analysis.
The upper four panels show the MEG$\pm 1$ (left) and
HEG$\pm 1$ (right) spectra from region R1. The lower four panels similarly show the 
spectra from region R9. As in the previous plots, the model is plotted with a red line. The nominal positions of S lines are marked at the top of each panel.}
\label{fig-hetg-s}
\end{figure}

Figure~\ref{fig-regions} shows the regions in the image of Cas~A selected for this
study. These regions have a morphology and brightness that allow
 the reliable measurement of line fluxes and centroids. They are spatially
narrow along the dispersion direction (north-south) and sufficiently isolated above the local, extended background emission 
to provide a clear line profile. Figure~\ref{fig-close-up} shows close-ups of these regions. 
For each region we concentrated on the
four strongest dispersed spectra: the MEG +1 and $-1$ orders and the HEG
+1 and $-1$ orders.  Because of their dominant emission we use the He-
and H-like transitions of Si and S lines for the analysis of
velocity and plasma structure in Cas~A. For typical SNR plasma
densities the He-like triplet of the Si and S line shows strong
forbidden ($f$) and resonance lines ($r$) and a comparatively weaker
intercombination line ($i$). We, therefore, jointly fit our four HETG
spectra (MEG$\pm 1$ and HEG$\pm 1$) with a model consisting of 
 4 Gaussians (representing 3 He-like lines and 1 H-like line for each element) 
 and a constant that accounts for the continuum level (for more details see
Appendix~A). Figure~\ref{fig-hetg-si} shows the four sets of Si spectra
from region  R1, the brightest of the 17 regions, and from region R9
which has the most prominent Si XIV line.
Figure~\ref{fig-hetg-s} shows similarly S spectra from these two regions. 
In the fit we allow only the centroid of the $r$-line to be a free parameter, 
while the other three centroids are tied to it, so the relative centroids are fixed, but the template is allowed to slide. We also assume that a single velocity is present in each Cas~A region, so all 4 Gaussians have frozen narrow widths. In other words, if there is no velocity structure in a region, the chosen Gaussian width would be the narrowest profile for that spatial structure. In addition, the flux ratio of the  $f$-line to the $i$-line 
(so-called R-ratio) for the Si triplet is tied to be 2.63, suitable for 
the very low density regime of the Cas~A plasma \citep[e.g.][]{porquet01}. Similarly, 
the $f$-line flux is tied to be 1.75 times the $i$-line flux for the S triplet \citep[e.g.][]{pradhan82}.

\begin{figure}
\centering
\includegraphics[height=12cm,angle=-90]{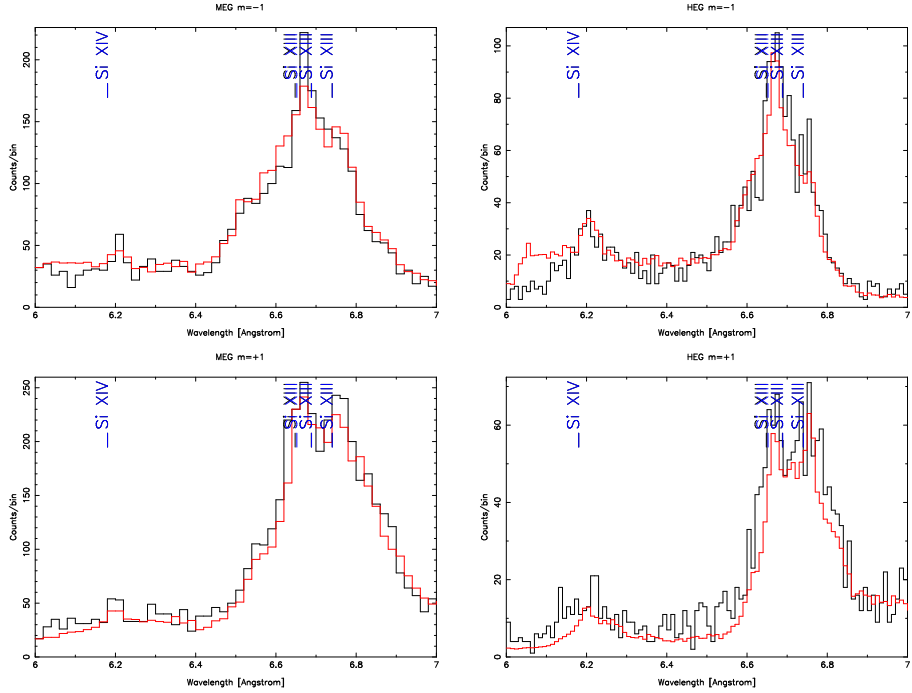}
\caption{Example of velocity-smearing in region R15 due to a velocity gradient 
in the dispersion direction (see also Fig.~\ref{fig-analysis-plots}(d)).
Note that for both the MEG and HEG spectra the $-1$ order (upper plots)
is narrower and more peaked than the +1 order (lower plots). As in previous plots, the model is plotted with a red line. The nominal positions of the Si lines are marked at the top of each panel.}
\label{fig-smeared}
\end{figure}

 To derive accurate Doppler shifts from measuring line centroids we used
 HEG and MEG spectra binned to 0.01 \AA\ and 0.02 \AA, respectively. Our 
Gaussian fits are not sensitive to measure velocity dispersion as
done in the case of E0102--072 \citep{flanagan04}, where  
topographical changes
in the dispersed image between positive and negative orders were used to
determine intrinsic bulk motions. The RMFs we use in the fitting of Cas~A HETG spectra only 
include broadening due to spatial
structure and do not include any further velocity structure degrees of
freedom in the quantitative fitting.  But we can qualitatively  
distinguish
regions with more or less velocity structure present. 
Among the line profiles from our 17 regions we find two qualitative types: 
1) a narrow profile, where the $r$ and $f$
lines are clearly resolved and shifted due to the region's bulk motion along the line of
sight, and 2) a smeared profile, where one of the grating orders 
appears broadened due to high-velocity motions within the region 
relative to its center of mass that causes distortion of the
dispersed images along the dispersion direction.  For each region the type of
the line profile is indicated in Table~\ref{tab-hetg}; Figure~\ref{fig-hetg-si} shows an example  of a double-peak profiles, and
an example of a clearly smeared spectral profiles is shown in Figure~\ref{fig-smeared}.

\subsection{Line Dynamics - Doppler Shift Measurement}

\begin{figure}
\centering
\includegraphics[height=12cm]{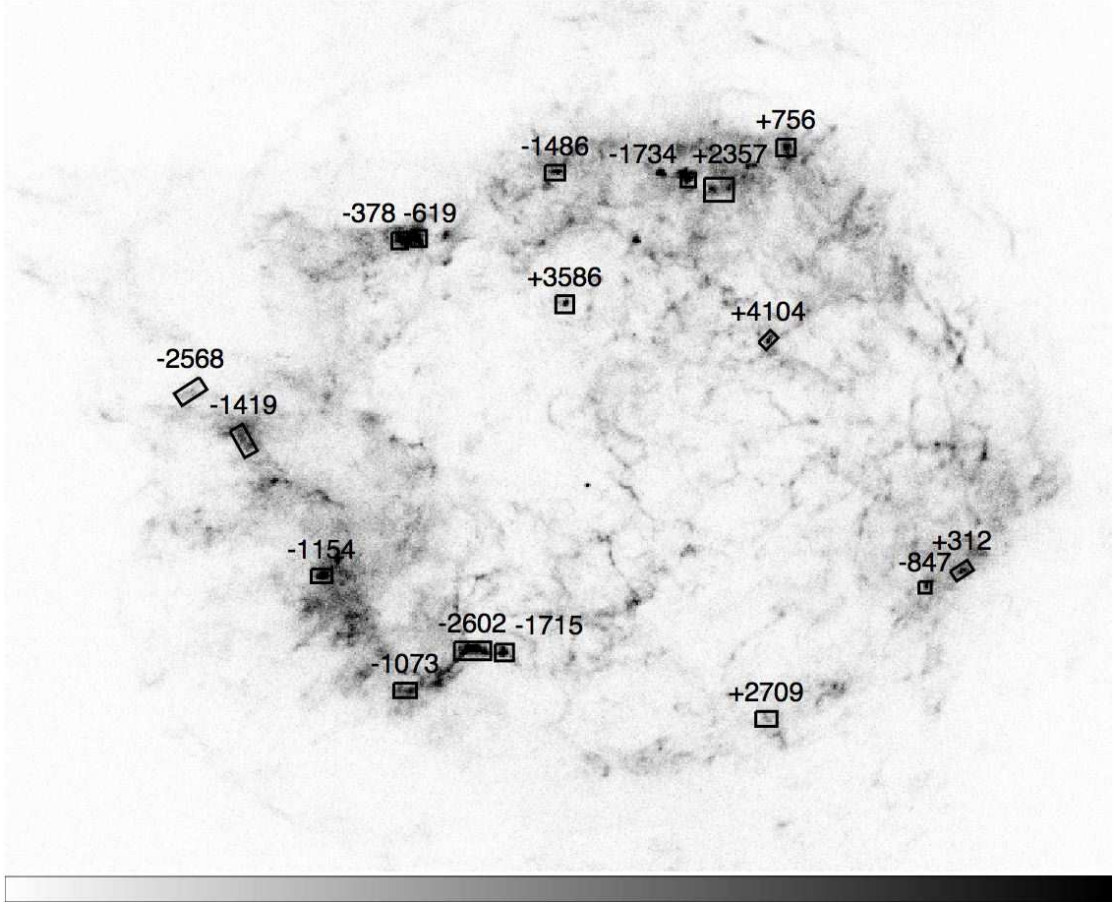}
\caption{Doppler velocity values in \kms\ for individual Cas~A regions.  The south-east SNR regions 
measured here are all blue-shifted, while extreme red-shifts are seen in the north-west of the SNR.}
\label{fig-doppler}
\end{figure}

\begin{figure}
\centering
\includegraphics[height=15cm]{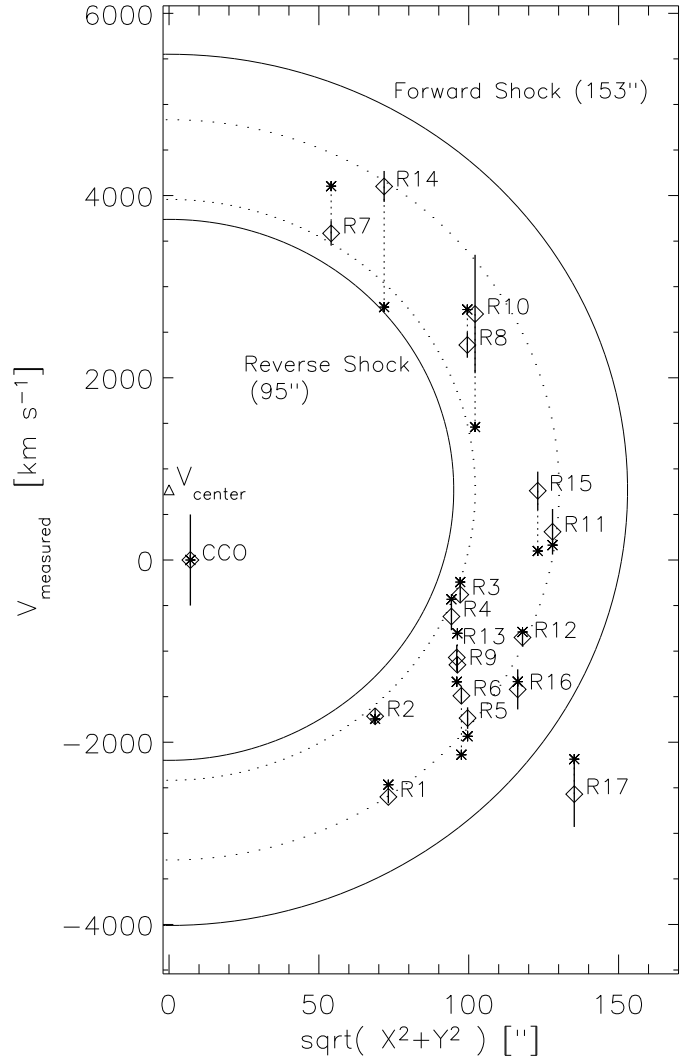}
\caption{3D Location and velocity of the regions.
The line-of-sight velocity, as measured by Si lines and S lines, is 
plotted vs.\ the 2D projected
distance on the sky from the nominal expansion center of \citet{reed95}. The Si velocity values (diamonds) are given with 90\%\ confidence error bars; the velocity values for S (stars) are given just for the best-fit value. An expansion rate of 0.19~\%\ per year relates the velocity and distance scales (see text).  The inner and outer solid lines indicate the locations of the reverse shock (95\arcsec)
and the forward shock (153\arcsec) in year 2000 and centered on the \citet{reed95}
velocity center (marked with triangle), ${\rm V_{center}}=+770$\kms.
Dotted lines at 102\arcsec\ and 130\arcsec\ are a guide to
show the shell in which our regions generally lie.
For reference, the location of the compact central object in Cas~A \citep[e.g.][]{chakrabarty00} is also shown; its line-of-sight velocity is unknown.}
\label{fig-projected}
\end{figure}

Results of Si line measurements and derived Doppler shift values are summarized in Table~\ref{tab-hetg}. The 
spatial distribution of Doppler velocities is shown in Figure~\ref{fig-doppler}. Derived values range between $-2500$\kms\ to $+4000$\kms. The measured velocity shifts for Si and S (not listed here) are similar (see Fig.~\ref{fig-projected}). This is not surprising since they arise from the same nucleosynthesis layer, and have been found to have the same spatial
distribution \citep{hwang00,willingale02}.  The uncertainties in derived velocities depend,
besides on the intrinsic energy resolution of the HETGS, on the number
of counts detected in the line and the errors associated with
estimates of the continuum contribution. For the Si line the statistical errors for the Doppler shift, based on fit confidence limits, range range within 200\kms\ for all regions except for regions R10 and R17 which have errors of 650\kms\ and 360\kms, respectively.

The HETGS derived velocities of our 17 Cas~A regions are combined
with their spatial location on the sky to graphically indicate
their 3D location and velocities, as shown in Figure~\ref{fig-projected}.
Our measured Doppler velocity is plotted on the y-axis and
the x-axis value is the 2D sky radial displacement of the region from the
expansion center of Cas~A on the sky given by \citet{reed95}.
A factor of $0.032\arcsec \pm 0.002\arcsec$ per \kms\ is used 
to relate velocity to spatial location by minimizing (by eye)
the shell width  needed to enclose most of the data points (dotted lines.)
The velocity center for the shells of +770\kms\ is taken from \citet{reed95}.  
For a distance to Cas~A of 3.4~kpc \citep{reed95} our factor corresponds to a
fractional expansion rate of $(0.19\pm 0.01)$~\%\ per year.
The forward shock location at 153\arcsec\ and the mean reverse shock location at 95\arcsec, as determined by \citet{gotthelf01}, is also indicated.

\subsection{Line Diagnostics - Flux Ratio Measurements}

One advantage of the high-resolution grating spectra of our Cas~A regions over
lower-resolution CCD data is that we can investigate the plasma state of individual regions using individual emission lines. The He-like K-shell
lines, like those of Si and S present in Cas~A, are the dominant ion
species for each element over a wide range of temperature
\citep[e.g.][]{paerels03}.  From the He-like triplet, the ratio of the forbidden
($f$) and resonance ($r$) lines is a useful diagnostic for electron
temperature \citep[e.g., the G-ratio $=(f+i)/r$,][]{porquet01},
especially since the lines are from the same ion, which
reduces dependence on the relative ionization fraction.
Lines from different ions of the same element are
also useful because they eliminate the impact of uncertainties
in abundance, so e.g. the ratio of the H-like to He-like Si
lines in conjuction with the G-ratio of the He-like lines 
can give an accurate measurement of the progression of plasma
ionization.

\begin{figure}
\centering
\includegraphics[height=10cm]{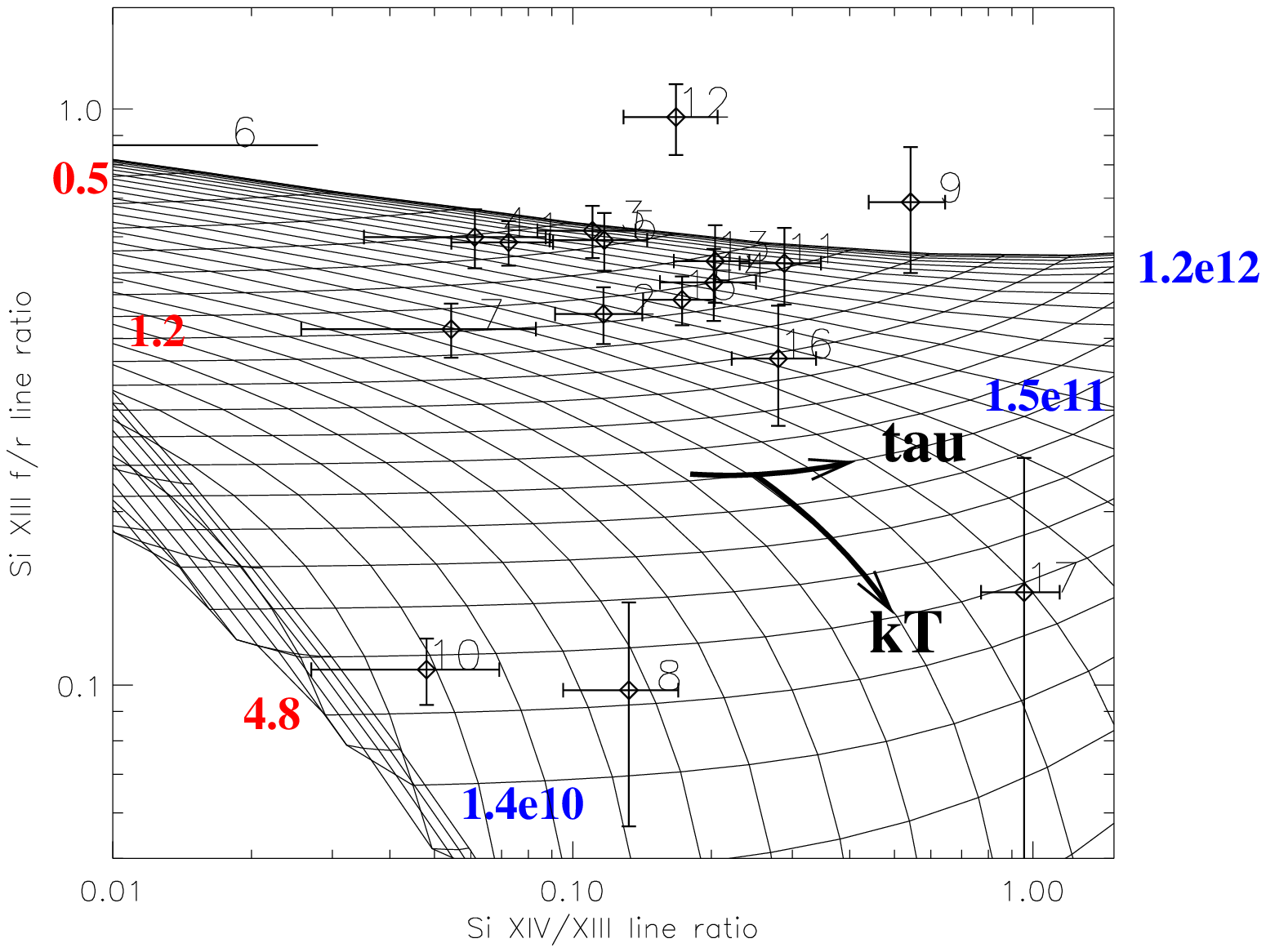}
\includegraphics[height=10cm]{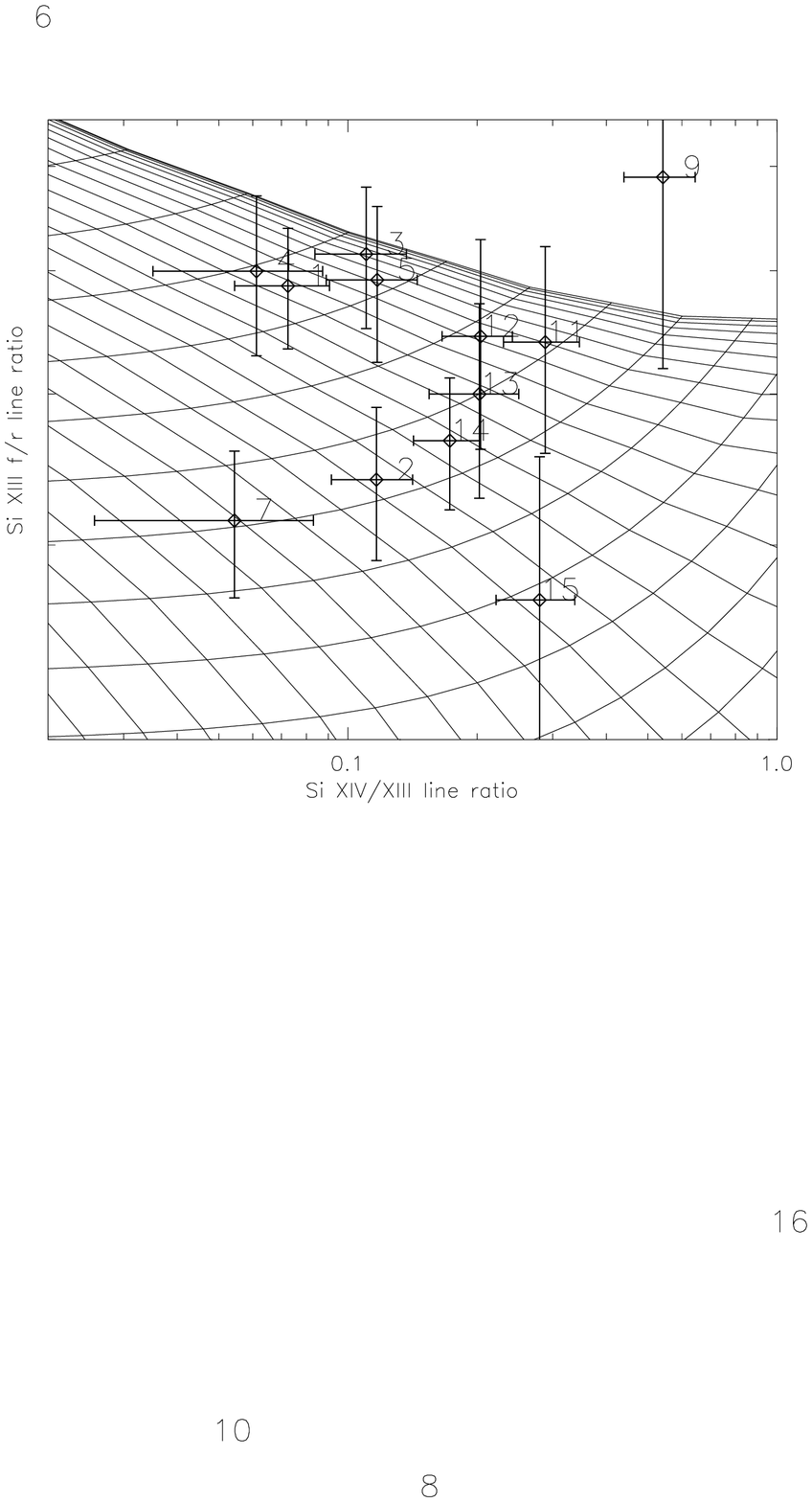}
\caption{Si line ratios overlaid onto a NEI grid. Upper panel shows the full range of values for all the regions (region R6 is located 
 off the graph), and the lower panel zooms in on the cluster of values in the upper region of the graph. 
Temperature, $kT_e$, increases along the lines which go generally
from upper-left to lower right. Ionization time, $\tau$,
increases along the lines which are nearly horizontal at
left and move to the upper-right ending at the high-$\tau$ limiting asymptote. A few $kT$ values are labeled in red, while some $\tau$ values are labeled in blue.}
\label{fig-line_ratios}
\end{figure}

To determine accurate Si line flux ratios from our data, we fixed the
modelled line locations based on our nominal binning and fits described above, and then
re-fitted HETG spectra using coarser-bin with modified errors.  This
procedure, described in Appendix~A, is less sensitive to differences
between the shape of the analysis RMF and the velocity-modified line shapes.
The resulting Si $f/r$ and Si XIV/XIII line ratios 
are listed in Table~\ref{tab-hetg}. We also calculated the expected line
ratios, employing the
non-equilibrium ionization collisional plasma model with variable
abundances, VNEI \citep{borkowski01} in XSPEC with vneivers version 1.1, which uses updated calculations of ionization fractions from \citet{mazzotta98}.
We then used these model grids to map our measured line
ratios to equivalent plasma temperature $kT_e$ and ionization
timescale $\tau = n_e t$.  Figure~\ref{fig-line_ratios} shows the distribution of
$kT_e$ and $\tau$ for the 17 Cas~A regions. The point for region R6
falls outside the displayed ratio range because its Si XIV/XIII
ratio has an extremely low value. The temperatures range
between 0.4 to 5~keV, with the majority of regions having a temperature
between 0.7 and 1.0~keV; the distribution of plasma temperature is
shown more clearly in Figure~\ref{fig-kT-tau}.  The ionization timescale ranges between
 \e{10} and 4\ee{11}\cms. The measured values for region R9 and R12 fall
off of the grid in Figure~\ref{fig-line_ratios} because of their
extremely high $f/r$ line ratios. To assign $kT_e$ and $\tau$ values
to region R9, we used the lower error limit which falls within the
NEI grid. For region R12, we used a value within a 20--30\% discrepancy of
the lower limit. Thus, the derived values (e.g., $n_e$, $t_{\rm shock}$)
for regions R6, R9 and R12 should be taken with some leeway.

\subsection{Zeroth-order Spectra - Density and Abundance Measurements}
\label{sec-zeroth}

In addition to the dispersed data, we also use the non-dispersed data
(from the central ACIS chip seen in Fig.~\ref{fig-casA}) to obtain 
information on the abundances and emission measure for the individual regions. Obtaining information on global plasma parameters such as electron density and elemental abundance requires comparing line to continuum in the spectra, and also  
knowing what are the contributions to the continuum. The dispersed spectra will have contribution from superimposed lines and continua from other image regions, although the superimposed lines would not have the right energies (see Appendix A for more details). Because of this confusion  dispersed data are not reliable for measuring line to continuum ratio.  Therefore, to determine these plasma parameters for each of the 17 regions, we fitted the non-dispersed 
zeroth-order spectra with a single VNEI plasma model \citep{borkowski01}, which allows for varying elemental abundances.  
In these fits we fixed the $kT_e$ and $\tau$ values for each region according to the values
derived from HETG line ratios. Cas~A spectra, even of isolated features, are very complex and four
spectral components have been identified \citep[e.g.][]{delaney04}. These four spectral components are rarely present at the same location and most of \citet{delaney04} knots show characteristics of a single
type, with some showing mixed characteristics. To avoid complications due
to varying column density ($N_{\rm H}$) and continuum levels we ignore the low-energy part of
the spectrum and consider only the range between 1.1 and 8.0~keV
encompassing K-shell lines of Mg, Si, S, Ar, Ca and Fe-K. An O-rich plasma is
often employed when describing Cas~A ejecta
\citep[e.g.][]{vink99,hughes00,laming03}, since optical observations showed 
that ejecta in Cas A are deficient in H and rich in O and O-burning
products \citep[e.g.][]{chevalier78}. 
  We, therefore, assume that O dominates the continuum emission, and that O provides many of the electrons instead of H and He, as in the typical solar abundance plasma. 
Thus, we fix the O abundance to be a factor of 1000 higher than the solar value. The results are, of course, not very sensitive to the exact O overabundance factor.  
We also fix $N_{\rm H}$ to 1.5\ee{22}\cm{-2}, which is found to be an average 
value across the SNR \citep[e.g.][]{vink96,willingale02}.  In the fit we allowed 
the normalization and the abundances of Mg, Si, S and Ca
to vary. The Ar line is not included in the
VNEI model (vneivers version 1.1), so we use a Gaussian to model the Ar emission. 
The prominent Fe-K XXV line is only present in regions R9 and R16, which is not surprising
since they are located in the area rich with Fe called the "Fe cloud"
\citep{hwang03}.

We tried different approaches for handling the red-shift parameter $z$ in
fitting these CCD data. We first froze the $z$ values to those derived from the HETGS data, but the resulting fits were not acceptable, showing a clear
mismatch between data and model line peaks.  These offsets are likely
ACIS gain uncertainties and other calibration errors in the CCD
response.  Reasonable fits were obtained by manually adjusting the 
$z$ values separately for each data set; the
values used varied between $-0.005$ and $+0.035$.

Table~\ref{tab-zo} lists parameters measured from the zeroth-order spectral fits
of the 17 regions.  
The first column lists the region emission volume $V_R$, which was taken to be a triaxial ellipsoid whose
2D-projected area corresponds to the spectral extraction region shown in Figure~\ref{fig-close-up}
with radii $a$ and $b$. 
For the third axis along the line of sight we take the average value of the two observed axes,
$ c = (a+b)/2 $; the volume of the region is then:
 
\begin{equation}
    V_R = {4\over 3}\pi \ a \ b \ c .
\end{equation}

\noindent
The second column lists the normalization factor (``norm'') used in the VNEI model, 
e.g., 

\begin{equation}
   X_{\rm norm} \ \equiv \  { {10^{-14}} \over {(4 \pi d^2)} } \ \int n_e n_H \ dV .
\end{equation}

\noindent Note that the tabulated norm here assumes that the model
oxygen abundance is set to 1000.
The rest of the columns list the abundance ratios with respect to
oxygen for the elements Mg, Si, S and Ca.  Using the equations  
given in Appendix~\ref{sec:vnei_to_phys} we derive and tabulate in
Table~\ref{tab-zo-param} some relevant physical parameters for the 
regions based on these fit results:  the region's
electron density, the total mass of plasma in the
region, the time-since-shocked for
the region $t_{\rm shock} = \tau / n_e $, in units of years, and
finally the fraction of oxygen by mass per region.


\section{Discussion}

\subsection{Doppler measurements in Cas A}

The measurement of Cas~A Doppler shifts in the X-ray band was first conducted with
the {\em Einstein X-ray Observatory} using the Focal Plane Crystal Spectrometer (FPCS) \citep{markert83}. The bulk velocities
of two regions of Cas~A, the SE and NW halves, were measured 
 using the line centroids
of the resolved Si and S triplets. These
observations found a velocity broadening and asymmetry in the X-ray emitting
material, with the NW region having more red-shifted emission, and the SE
region of the remnant having more blue-shifted
emission. \citet{markert83} suggested that the asymmetry could be the
result of an inclined ring-like distribution of Cas~A material,
possibly influenced by the distribution of the mass-loss material of
the progenitor. Our HETGS spectra of individual Cas A
filaments reconfirms this global asymmetry trend. The SE regions 
of the SNR appear to be mostly blue-shifted and the regions
in the NW have the extreme red-shifted values. 
The asymmetry was also found with moderate spectral resolution of {\em ASCA}
\citep{holt94}. The {\em ASCA} observations provided a velocity map on
the spatial scale of 1\arcmin\ and were derived using the Si line
centroids at the CCD resolution.

Doppler velocity maps with a much finer spatial resolution were
produced using {\em Chandra} \citep{hwang01} and {\em XMM-Newton}
observations \citep{willingale02}.  \citet{hwang01} used the Si XIII
resonance (1.865~keV) and Si XIV Ly$\alpha$ (2.006~keV) line centroids to
derive the velocity shifts on a 4\arcsec\ spatial scale. While most of
the SNR showed velocities between $-1500$ and $+1500$\kms,
extreme velocities of $-6000$\kms\ were found in the SE region,
including the region of our R1, R2, R9 and R13 for which HETGS data
imply velocities between $-2500$ and $-1000$\kms. 
This discrepancy is not too
surprising. Although \citet{hwang01} argued that they can ignore the
ionization effects on the Si He$\alpha$ centroid to a reasonable
approximation, they note that high spectral resolution measurements
are more desirable to measure the energy shifts directly from resolved
rather than blended lines. Indeed, their {\em Chandra} data do not
resolve the Si XIII triplet, and even the Si XIV and Si XIII Ly$\alpha$ lines
are not fully resolved, introducing some level of
uncertainty in line centroid measurement. 
The most recent results from the 1~Ms Very Large
Project (VLP) {\em Chandra} data show values more consistent with our HETGS data 
\citep{stage05}, but even here absolute gain calibration accuracies for the
CCD may only be good to 0.5\%\ or 1500\kms.

{\em XMM} Doppler maps of Si, S and Fe lines by \citet{willingale02}
show a smaller range in velocity than the Si velocity map of
\citet{hwang01} and, thus, values that are at a glance closer to our values in
Table~\ref{tab-hetg}. The {\em XMM} data have been binned to a $20\arcsec
\times 20\arcsec$ spatial grid and the spectra were fit with two
thermal NEI components representing the ejecta and the shocked
component, each with a separate energy shift. The Doppler shift values
derived in this way will depend on the ability of the NEI spectral
model to predict the line blends combined with the uncertainties in
the gain calibration of the detectors. In comparison to our results,
it is obvious that some of the errors in their Doppler shift values
are produced because of the spatial averaging over a variety of features with
very different velocities. For example, in the NW region the Doppler
velocity map of \citet{willingale02} has a smooth distribution of
red-shifted values with a 1000\kms\ or so dispersion, whereas our
regions R5 and R6 in that part of the SNR show significantly
blue-shifted values of around $-1500$\kms, as well as red-shifted
velocities with 1000\kms\ difference between regions R8 and
R15. \citet{willingale02} find that the velocity patterns for S are
very similar to those for Si; our measured Si and S velocities agree
with this. The Fe-K velocities, however, exhibit higher velocities than
those of Si and S; this has also been found with the 1~Ms VLP
{\em Chandra} observations of Cas~A
\citep{stage05}. Note that our two regions showing Fe-K emission, R9
and R16, are located in the middle (R9) and outside (R16) the range of
other regions (see Fig.~\ref{fig-projected}).  Unfortunately, our HETGS data do not have enough
counts in the Fe-K band to measure shifts in the Fe energy; deeper
HETGS observations might yield Fe-K velocity measurements with
500\kms\ errors, provided the required narrow features are present.

Probably the best demonstration of the magnitude of the difference
between the CCD and HETGS measured Doppler shifts is given by
comparing the estimated shock expansion rate in
Cas~A. \citet{delaney03} derived forward shock expansion measurements 
of 0.21\%\ per year using transverse velocity
measurements of Cas~A knots using {\em Chandra} CCD data. \citet{delaney04}
 compared their transverse velocities (3100--3900\kms) with that of
 \citet{willingale02} (1000--1500\kms) and \citet{hwang01} (2000-3000\kms) 
obtained from Doppler measurements using CCD spectra, 
 and graciously ascribe mismatch to possible projection
effects of the asymmetric remnant.
However, our Doppler measurements imply an ejecta expansion of
0.19\%\ per year, consistent with the \citet{delaney03} and
\citet{delaney04} value.  This
supports their suggestion that there might be a dynamic coupling
between forward shock and ejecta, and they are both part of one
homologously expanding structure. Note that \citet{vink98} derived
 an expansion rate of 0.2\%\ per year by comparing the {\em ROSAT} and
{\em Einstein} X-ray images from two different epochs.
 
The region R17 is the only region outside our
reverse-shock/forward-shock range and its location at a large radius
from the expansion center puts it outside the nominal forward shock 3D
radius as seen in Figure~\ref{fig-projected}. This is also a
characteristic of the optical FMKs, which are found mostly in the NE
part of the remnant, with many of them located along the jet. Our
region R17 is located in the base of the jet region (see Fig.~\ref{fig-regions}), and its velocity indicates that it indeed might be part of the jet feature.


\subsection{Plasma properties}

\begin{figure}
\centering
\includegraphics[height=10cm]{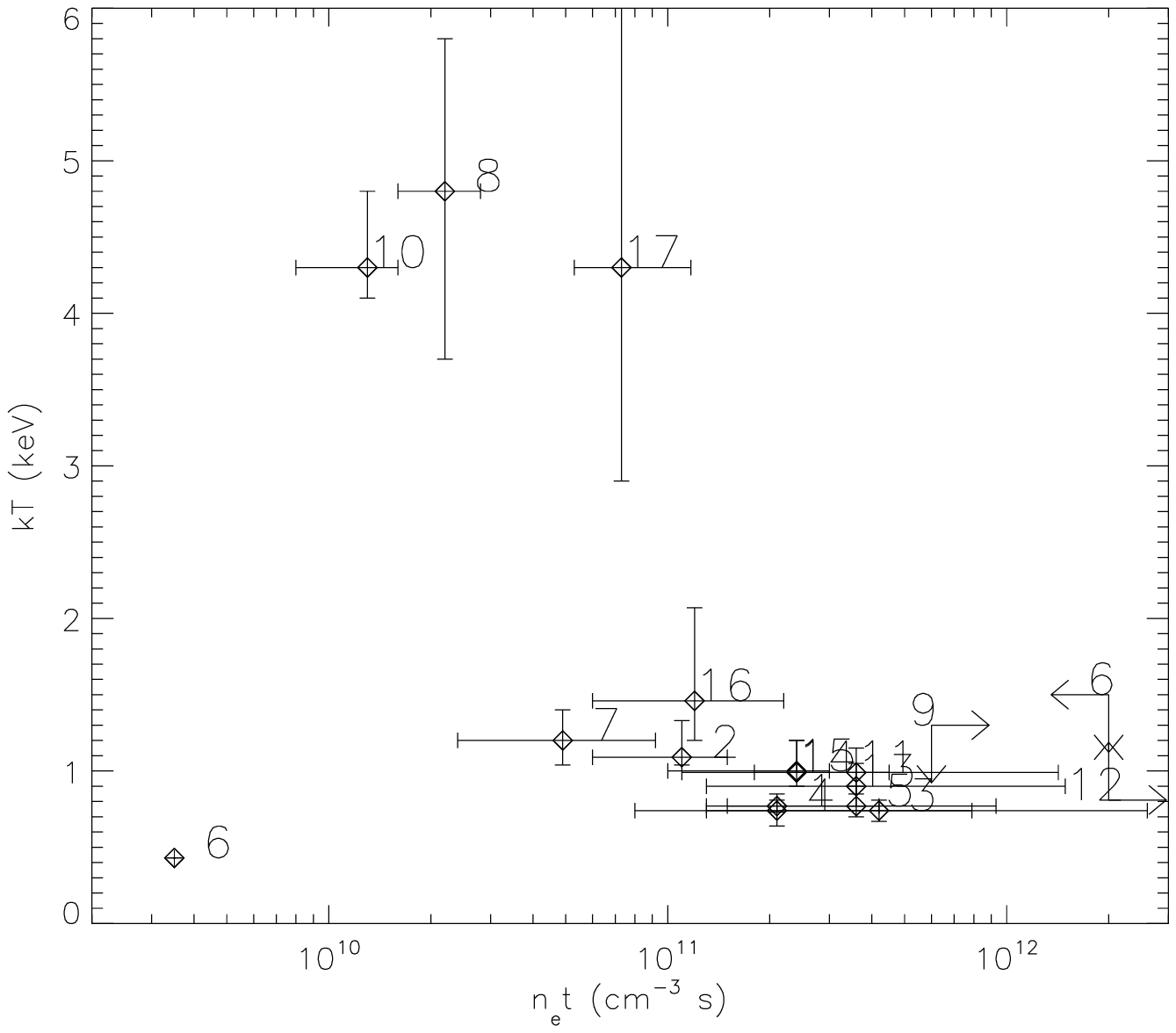}
\includegraphics[height=10cm]{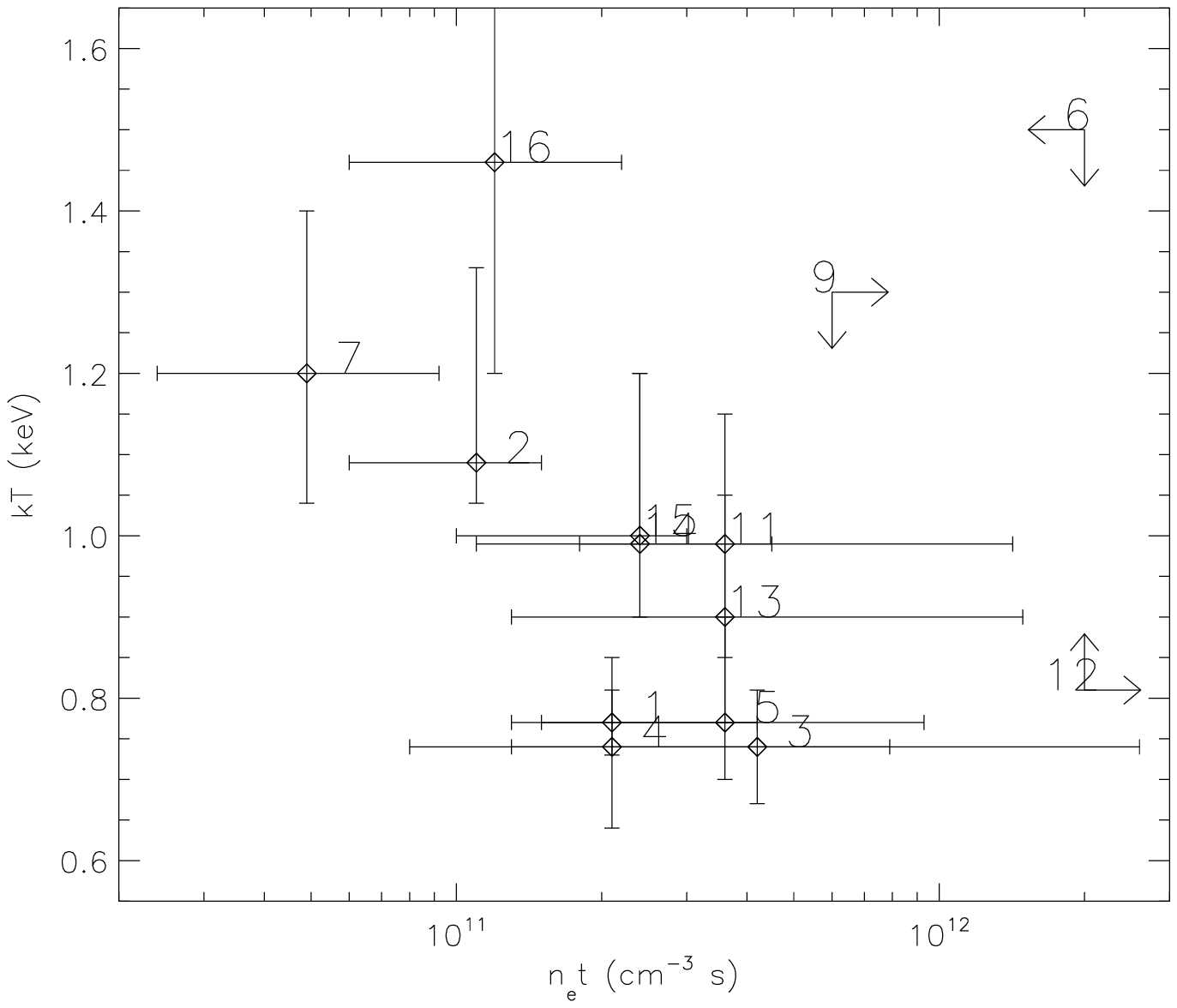}
\caption{Plot of plasma temperature $kT_e$ vs. ionization timescale $\tau=n_e t$ for 17 Cas~A regions. The upper panel shows the full range of values, and the lower panel zooms in on the cluster of values. Three regions, R8, R10 and R17, show different properties from the rest. Region R6 is represented with two points, representing the mean value (lower left corner in the upper panel) and the upper limits (lower right corner in the upper panel).}
\label{fig-kT-tau}
\end{figure}

Beside the Doppler shifts, plasma temperature and ionization timescale
are the two other plasma parameters derived directly from our HETGS
analysis. Early observations of Cas~A in X-rays identified two distinct
plasma components --- the cold component with temperature around
0.6~keV, and the hot component with temperature up to 4~keV
\citep[e.g.][]{becker79}, and they have been confirmed with newer observations 
\citep[e.g.][]{vink96,willingale02}. The cooler component is
associated with the reverse shock traveling through the expanding ejecta, and the hotter
component is associated with the forward shock propagating into the circumstellar material. Plasma temperatures
derived from our HETGS data have values mostly around 1~keV.
Therefore, most of our regions have been heated up by a reverse shock
propagating inward into the supernova ejecta. There is no significant
pattern to the variation in the temperature across the SNR, as shown in
Figure~\ref{fig-kT-tau}. The exceptions are regions R8, R10 and 
R17 which do have temperatures of 4--5~keV,  
several times higher than most other regions. Regions R8 and R10, which also share similar velocity, could, therefore,  be associated  
with circumstellar material and the forward shock. 
Region R17 could be an X-ray counterpart of the optical FMKs found in Cas~A jet. 
\citet{willingale02} derived a map of ionization timescale for the 
cool plasma component in Cas~A, which shows some variation across 
the surface of the SNR, but most of the SNR has $\tau$ values larger 
than \e{11} \cms. For our 17 regions we also find little variation, 
most of the regions have $\tau$ around a few \e{11}\cms. The exception are, again, 
only regions R8 and R10, as shown in Figure~\ref{fig-kT-tau}, which have ionization timescale up to an order of magnitude smaller.

\begin{figure}
\centering
\includegraphics[height=12cm,angle=-90]{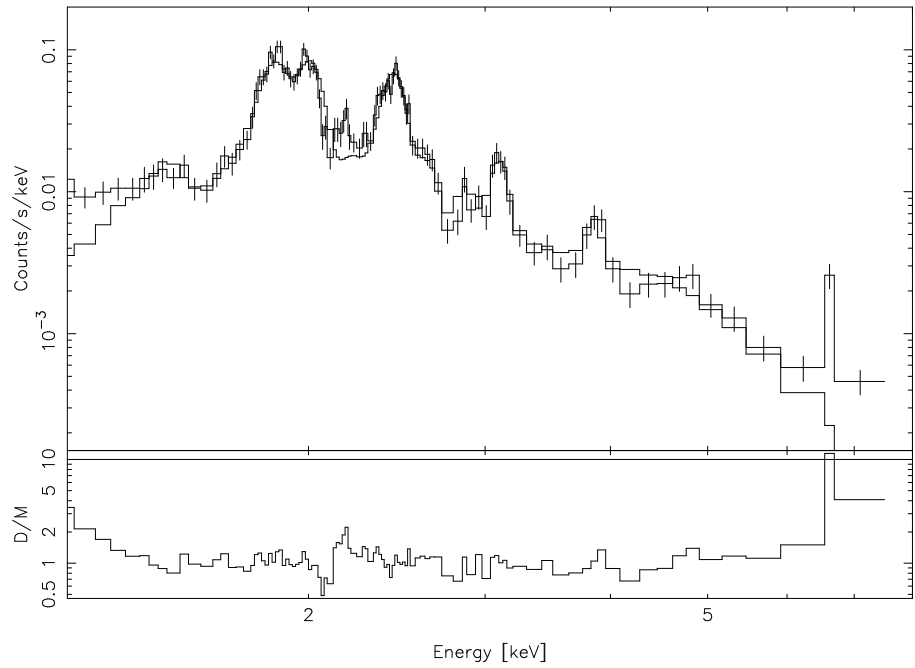}
\caption{The zeroth-order spectrum from region R9. The spectrum
  shows a good fit to the Mg, Si S, Ar and Ca lines and to the continuum part of the spectrum between 1.2 and 6~keV; Fe lines were not included in the fit and
the part of the spectrum below 1.1~keV is not used in the fit. }
\label{fig-r9-zo}
\end{figure}

Information on density and abundances for the 17 Cas~A regions is
derived in conjunction with the zeroth-order data. The zeroth-order spectra
 are reasonably fit with a single VNEI model with fixed temperature and ionization timescale, certainly well enough for our
primary purposes here to establish relative abundances. An example spectrum
 is given in Figure~\ref{fig-r9-zo} for region
R9. In the 1.1 and 8~keV range the spectrum shows a weak Mg lines, strong Si, S, Ar and Ca lines, and a weaker Fe-K line, and shows reasonable agreement between the data and the
model for the continuum part of the spectrum. Note that a low-energy
``up-turn'' is also present due to emission from Fe-L lines and this part of
the spectrum, below 1.1~keV, was not used in the fit.  The Si XIII Ly$\beta$ line at 
2.18~keV is generally
under-fit in the spectra; this is likely due to several calibration issues each
at the 10-20\%\ level (e.g., Ir contamination is not included in the
HRMA model, there are possible zeroth-order HETG calibration errors,
CCD gain offsets are coupled with steep ARFs.)  Future work using the
higher-count Cas~A VLP data set and including these calibration
effects could usefully extract the He-like Ly$\beta$ line flux allowing
associated diagnostic ratios \citep[][]{porquet01}
and possibly indicating charge-exchange processes \citep{pepino04}.
  
\begin{figure}
\centering
\includegraphics[height=15cm]{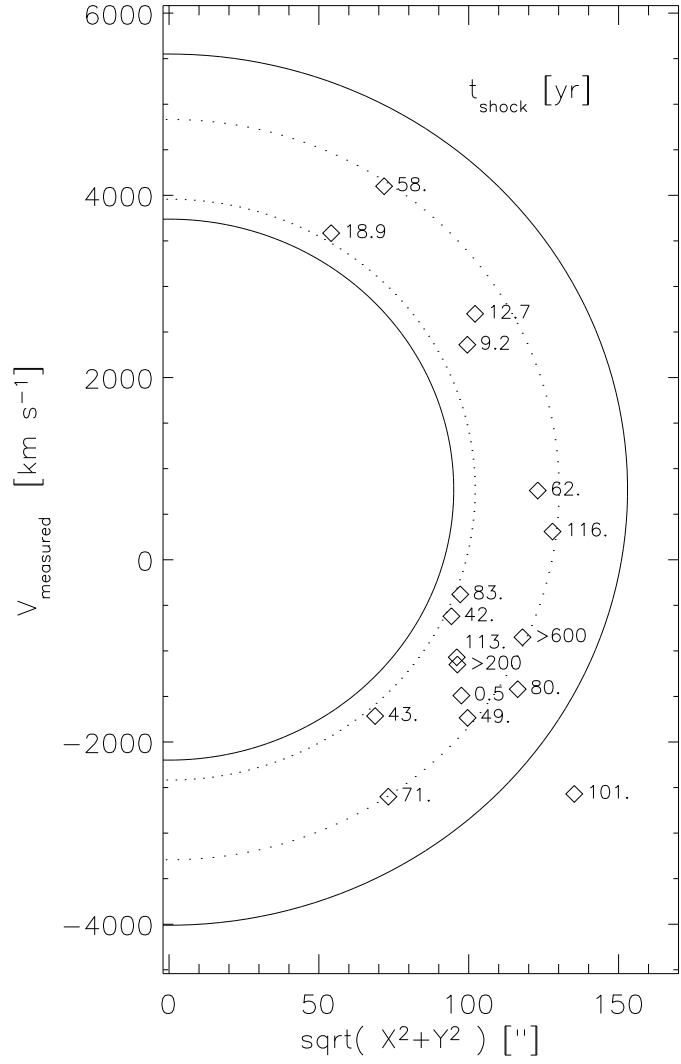}
\caption{  The time-since-shocked in years is indicated for each region in a similar plot to Fig.~\ref{fig-projected}.
The red-shifted regions appear to have been shocked more recently compared to the regions on the front side which have longer $t_{\rm shock}$ values; the low Si XIV region, R6, with 0.5~yr is an exception.}   
\label{fig-tshock}
\end{figure}

Electron density for each of the regions, derived as described in Appendix~B, 
is listed in Table~\ref{tab-zo-param}. The $n_e$
ranges from around 20 to 200\cm{-3} and it seems to have higher values
for the blue-shifted regions. A factor of 5 density difference has been
suggested between the front and the back of Cas~A
\citep{reed95}. The time since individual Cas~A regions have been
shocked, given by $t_{\rm shock}=\tau/ n_e$, is also listed in
Table~\ref{tab-zo-param}. These times vary significantly from region to
region and it seems that the red-shifted regions have been shocked
more recently compared to the regions on the front side which have generally
larger $t_{\rm shock}$ values (regions R6, R9 and R12 have extreme values 
that should be taken with caution), as
shown in Figure~\ref{fig-tshock}. However, these values could be the
result of density differences between the front and the back side of
the SNR.  We do not find any correlation between $kT_e$ and $t_{\rm shock}$ or 
$n_e$ and $\tau$ that would indicate a possible electron-ion equilibration 
occurring in our 17 regions.

The other parameters listed in Table~\ref{tab-zo-param} are the total
mass and the fraction of O mass per region. These values should be taken with caution, 
because we assumed in our fit that continuum is dominated by O.  Note also that our total mass
estimate does not include Fe or Ar since they are not included in our
NEI model, so the $M_{\rm total}$ values could be somewhat modified when these
are included. The O mass fraction ranges from 0.82 to 0.97, thus
confirming that, under our assumptions, O is the dominant element in
these Cas~A ejecta features.
Cas~A is an O-rich SNR, as indicated by optical observations, and this
leads to the further assumption that perhaps there is also a lot of O
in the regions between the bright Si knots that we see. For example,
 \citet{laming03} estimate that a density enhancement of around 2
coupled with the presence of the Si-Ca metals would allow these
knots to stand out even though surrounded with a pure oxygen plasma.
This may be another
reason that the O-band dispersed image appears so smeared out and that we do not detect 
clear O-lines in our HETGS data: there
is O emission everywhere within the shocked ejecta region so we do not
see discrete O filaments or clumps like those in the Si and S
band images. 

 In order to test the suggestion by \citet{laming03} about the wide-spread presence of O emission, we make a few simple estimates. 
 Table~\ref{tab-zo-param} shows the mass fraction of oxygen to be 0.82 in region R1. Under 
 the assumption that the entire SNR volume within a radius of 100\arcsec\ and 130\arcsec, for example,  is filled with oxygen of the density projected for R1, the total mass of oxygen would amount to 85\msol .  Earlier studies imply a mass of 4\msol\ for the entire ejecta at most \citep[e.g.][]{vink96}. Thus, our results seem to either overestimate the oxygen density or the density of that particular region is not representative for such  a large volume. Even if we further assume that the density in R1 is enhanced with respect to the ambient medium
by the factor of 2 to 5 \citep{reed95}, we would wind up with 40 to 16\msol\ of oxygen 
under uniform conditions. Introducing a density gradient to the ejecta, such as a rise in density towards the center of the remnant, gives an increase of at least
another factor of 2. Therefore, such simple estimates do not seem to support 
 the \citet{laming03} assumption. Alternatively, \citet{willingale03} derived a filling fraction of 0.009 for the ejecta component by assuming a pressure equilibrium between hot and cool plasma components, which lead to an electron density range of 40--90\cm{-3}, similar to our values.   
 Adding such a factor to our estimate above would then allocate much of the metals and oxygen into spaghetti-like filaments
requiring a total mass of order 1\msol.
This value represents a more reasonable contribution to a total 
oxygen mass of 2.6\msol\ suggested by \citet{vink96}.

\section{Summary}

High-resolution HETGS observations from the young SNR Cas A yield unprecedented kinematical X-ray results for a few bright SNR regions. These observations 
show that high-resolution X-ray spectroscopy is catching up with
 that in optical, IR and UV bands in its ability to measure velocities and add a third
dimension to the data. Unambiguous Doppler shifts are derived for these selected regions, with the SE region of the SNR showing mostly blue-shifted values reaching up to $-2500$\kms, and the NW side of the SNR having extreme red-shifted values with up $+4000$\kms. This global asymmetry is consistent with previous lower spatial or spectral resolution X-ray observations.  From our Doppler measurements we derive 
ejecta expansion of 0.19\%\ per year, supporting suggestion of \citet{delaney04} 
 that there might be a dynamic coupling between the forward shock, that is expanding at the same rate, and the ejecta. 

Plasma diagnostics using resolved Si He-like triplet lines
and Si H-like line shows that most of the selected regions in Cas~A have temperatures around 1~keV, consistent with reverse-shocked ejecta. However, two regions, R8 and R10, with significantly different temperature of  $\sim4$~keV, might be part of the circumstellar material. Similarly, the ionization age does not vary considerably across the remnant, except for the two above mentioned regions, which have an order of magnitude lower ionization timescale. One of the regions, R17, is located outside the forward shock boundary and also has a high plasma temperature, both suggesting that this region is part of the Cas~A NE jet feature.

The analysis of HETGS data was complemented by the non-dispersed CCD ACIS spectra from the same
 observation, which allowed us to derive the electron density of the
X-ray emitting ejecta and the elemental abundances of Mg, Si, S and Ar in our 17 regions.   The electron density varies from 20 to 200\cm{-3} and does not show any correlation with  
ionization timescale. The derived elemental abundances of
Mg, Si, S and Ca are consistent with O being the dominant element in the Cas~A plasma. 

\acknowledgments

We thank John Houck and John Davis for contributions in the planning of the Cas A HETGS
observation, and Glenn Allen, Mike Stage, Kathy Flanagan, Tracy
DeLaney, Dick Edgar, and Dan Patnaude for useful discussions.
Support for this work was
provided by NASA through the Smithsonian Astrophysical Observatory
(SAO) contract SV3-73016 to MIT for support of the Chandra X-Ray Center and Science
Instruments, operated by SAO for and on behalf of NASA under contract NAS8-03060.


\appendix

\section{Filament Analysis}
\label{sec:fil_analysis}

\subsection{Filament Analysis Scheme}

The ``filament analysis'' we used for Cas~A HETGS data 
is very similar to the standard analysis of dispersed grating data
from a point source,
in that it produces a one-dimensional PHA file
with associated ARF and RMF for each grating and order.
The two main additions to the standard processing that adapt it to extended,
filament-like sources are described briefly below.

\begin{figure}
\centering
\includegraphics[height=8cm]{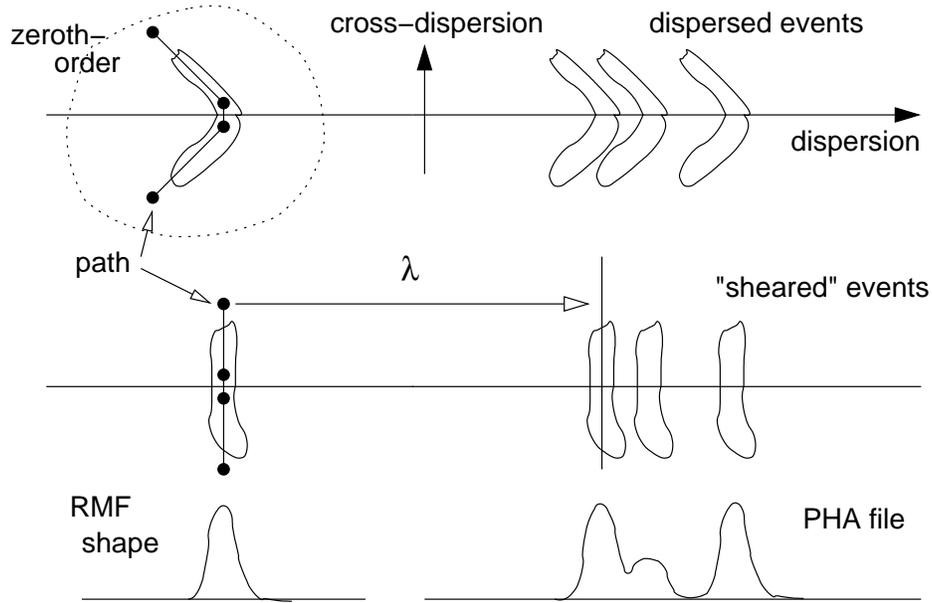}
\caption{Filament analysis schematic.  A path is defined and used to straighten a filament-like feature seen in zeroth-order. A "shearing" process is carried out to translate each event location in the dispersion direction by an amount given by its cross-dispersion coordinate and the path. An effective RMF and corresponding PHA file are created and can be used in data analysis.}
\label{fig-analysis}
\end{figure}

First, adjustments to the event locations are applied to effectively straighten the
filament-like source perpendicular to dispersion while retaining wavelength accuracy; specifically, the following
steps are carried out. The shape of the source in zeroth-order is manually traced by a piecewise-linear path that is saved as a set of vertices. The vertex locations are then transformed into coordinates aligned with the dispersion and cross-dispersion direction of the particular spectra (HEG or MEG) being extracted, see top diagram in Figure~\ref{fig-analysis}.
Each event, in both the zeroth-order and the dispersed orders, is then translated
along the grating dispersion direction by an amount equal and opposite to the
path offset at that same cross-dispersion location. This "shearing" causes the zeroth-order and any dispersed line-images of the feature of interest (FOI) to become narrower
along the dispersion coordinate, see middle diagram of Figure~\ref{fig-analysis}.
A PHA file of counts per dispersion bin is then created in the usual way by
projecting the events along the dispersion axis, see bottom diagram Figure~\ref{fig-analysis}.
Because the dispersed features are effectively narrowed, the ability to detect
and resolve discrete lines from the feature is improved.

Second, a companion response matrix, RMF, is
created based on the observed, sheared
zeroth-order events.  For each wavelength in the RMF,
a subset of events are selected using the available non-dispersive
energy reported by the CCD detector and the expected
response histogram is created and stored with appropriate offsets in the RMF.
Operationally this is similar to the effect of the {\tt rgsxsrc}
convolution model available in XSPEC\footnote{~~
http://heasarc.gsfc.nasa.gov/docs/xanadu/xspec/manual/XSmodelRgsxsrc.html  }
for use with XMM-Newton Reflection Grating Spectrometer data.

This is just one approach to extended source grating analysis \citep{dewey02}
and has the advantage of producing familiar PHA files which can be analyzed with
standard software like ISIS \citep{houck02}.   Note, however,
that it is fundamentally an approximation to a
fully multi-dimensional, spatial-spectral method
and so will necessarily have limitations; some of these are implicit in the
considerations and techniques described in the following sections.

\subsection{Treatment of Spectral Continua}

In addition to line emission from the narrow FOI,
there is also continuum emission from the feature, as well as line and
continuum emission from other parts of the extended source. These give
rise to a continuum component in the observed PHA spectrum which
we discuss and estimate here.

The observed count rate at a given location on the
detector is given by a 3D integral over the source flux as a function of
wavelength and position on the sky \citep[see eq.(49) in][]{davis01}.  
This integral includes the grating response function, which has the property that the
location of a dispersed photon includes a continuous dependence on the
photon's wavelength.  Thus, for a given position on the detector there
is a set of sky locations and corresponding wavelengths which all
contribute detected counts in this same location.  This is in contrast
to the point-source case where the point source introduces a delta
function in the integral and preserves a one to one mapping of
source wavelength to dispersed location in each grating-order.
The continuum counts seen within a bin in the PHA distribution file will
then consist of the sum over the spatial regions along the dispersion
axis weighted by their flux at the grating-equation-allowed
wavelengths.  Equivalently, the resulting spectrum includes not only
the true continuum from the FOI, but the overlapping, shifted spectra
from all regions along the dispersion axis.

In practice, the integral may be effectively truncated, e.g., if the spatial
extent of the source is moderate.  In addition, even for very
extended sources, the inherent energy resolution of the detector can 
be used to truncate the integral to a limited range in pulse-height,
reducing the artificial continuum level.  In this case, the
observed continuum will be of order a factor of $R_g / R_{\rm CCD}$ greater than the 
true FOI continuum level, where $R_g$ and $R_{\rm CCD}$ are the
effective resolving powers of the grating and order-sorting detector, respectively. Note that $R_g$ for a given feature varies like $1/E$
and $R_{\rm CCD}$ varies roughly like $\sqrt E$, where $E$ is the energy; thus,  the $R_g/R_{\rm CCD}$ ratio varies approximately like $E^{-1.5}$  (or $\lambda^{1.5}$). This is a small variation, of order $\pm$12\% over the 6--7 \AA\ range for Si lines.

\begin{figure}
\centering
\includegraphics[height=8cm]{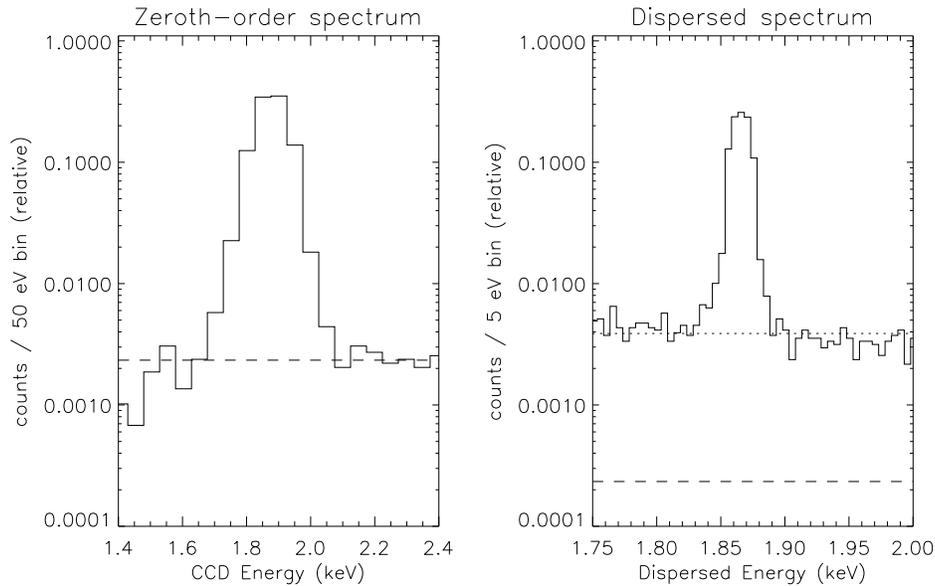}
\caption{ Elevated continuum level in dispersed spectrum. The zeroth-order and MEG-first order
spectra are shown for a simple MARX simulation of a monochromatic source embedded in a
large region of continuum.  The spectra are normalized to unit flux-in-the-line so that
the equivalent widths, EW, of the lines serve as a measure of the continuum level.
The EW for zeroth-order (left, dashed line) is 21.4~keV, whereas the dispersed spectrum has
a reduced EW of 1.29~keV (right, dotted).
Hence, relative to the line flux, the dispersed spectrum has
an artificially higher continuum level due to overlapping continua
from other locations along the dispersion axis.}   
\label{fig-analysis-cont}
\end{figure}

As a demonstration of this, a simple MARX simulation was made consisting
of a $\approx4.0\arcsec\times10\arcsec$ rectangle emitting at 1.865 keV
embedded in a disk of emission 100\arcsec\
in radius having a uniform spectrum from 1.2 to 2.8 keV.
Figure~\ref{fig-analysis-cont} shows the CCD spectrum extracted from the zeroth-order
(e.g., the rectangular region) and the spectrum extracted
from the MEG dispersed first order.  The effective grating resolving power here is
$R_g \approx 75$ based on the source full width.
For the order-sorting effective resolving power, $R_{\rm CCD}$,
the dispersed extraction was performed with a wide pulse-height
selection including $\pm 0.15\lambda$, giving
$R_{\rm CCD} \approx 3.3$, based on the energy full width.
Together these give $R_g / R_{\rm CCD} \approx 23.$
In the simulation the equivalent width, EW, of the line is 21.4 keV for the
zeroth-order case, and 1.29 keV in the dispersed case;
the continuum level is, therefore, $\approx 17$ times larger
in the dispersed data set, which is of the order expected
from the simple estimate.

\subsection{Velocity Effects and Fitting}

The features we observe in Cas~A show Doppler shifts of up to several
thousand \kms.  For a simple bulk motion of the emitting region this
just produces an overall Doppler shift to the line wavelength which
can be readily measured using the
filament analysis products.  It is also
possible that there are velocity variations within the filament itself
and this can introduce some complications, especially when carrying out
standard $\chi^2$-driven fitting.

\begin{figure}
\centering
\includegraphics[height=15cm]{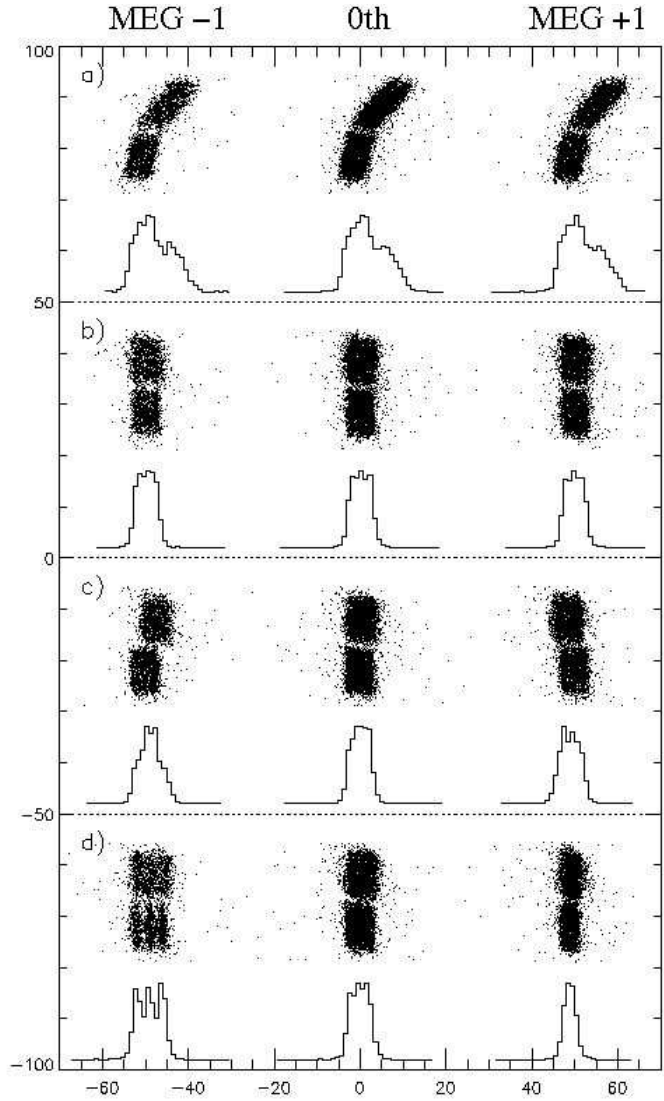}
\caption{Velocity effects on dispersed line shapes.
Each of the four panels here is from a simple MARX simulation and
filament analysis, showing close-ups of the minus order (MEG $-1$), zeroth-order (0th),
and plus-order (MEG +1) with event scatter-plots and 1D histograms.
The top panel (a) shows the original,
simulated filament without shearing applied.  In (b) all events have been
sheared using a simple path; the width of the feature has been narrowed
to about 5 pixels wide.  In (c) the upper-half of the feature is
emitting at a wavelength blue-shifted by 1000\kms\ with respect to the lower half.
The zeroth-order is unaffected, but the first orders are blurred equally  by this velocity variation in the cross-dispersion direction.
In (d) the source has a velocity variation along the dispersion direction
with blue-shift values of: 0, -500, and -1000\kms.  In this case the effect
is to broaden the minus order (showing the three individual line sources
used to simulate the wide filament) and narrow the plus order to the point
that it is actually narrower than the zeroth-order itself; the zeroth-order is unaffected. }
\label{fig-analysis-plots}
\end{figure}

In order to study these velocity effects, simple MARX simulations were
carried out using a filament source (resembling our region R1) consisting of 
two sets, upper and lower,  of three closely
spaced (by precisely 1\arcsec) parallel line sources. The two sets were
tilted and offset to produce a wide, kinky filament,
 shown in panel (a) in Figure~\ref{fig-analysis-plots}.  In the MEG simulations
of panels (a) and (b) all the 6 line sources making up the filament all have the same
wavelength.  These two panels show how the shearing of the filament
analysis improves the resolution of the dispersed spectra.

  In simulation (c) the upper set of three lines have all been
blue-shifted by 1000\kms\ demonstrating the effect of velocity variation
along the filament.  This causes the dispersed projections to be additionally blurred, having a FWHM larger than the (unaffected) zeroth-order.  A similar result
would arise for the case of turbulent velocity broadening where there would
be a range of velocities in all parts of the emitting feature.

  The simulation in panel (d) is perhaps the most pathological: here the
velocity of the filament varies across the filament, along the dispersion
direction.  The zeroth-order continues to be unaffected by
these velocity effects, and now not only do the the dispersed order
projections differ from the zeroth-order, but they are different
from each other as well \citep[see HETGS observations of SNR E0102--072,][]{flanagan04}.

  The effect of these velocity produced changes in the dispersed
line profiles is that the RMF created from the zeroth-order is
not a good match when there is velocity structure in the feature.
For purposes
of fitting the line location, the generally peaked nature of both the 
RMF and the dispersed line peaks leads to reasonable centroid
fitting and confidence ranges; however, the formal $\chi^2$ values
can be large.

\begin{figure}
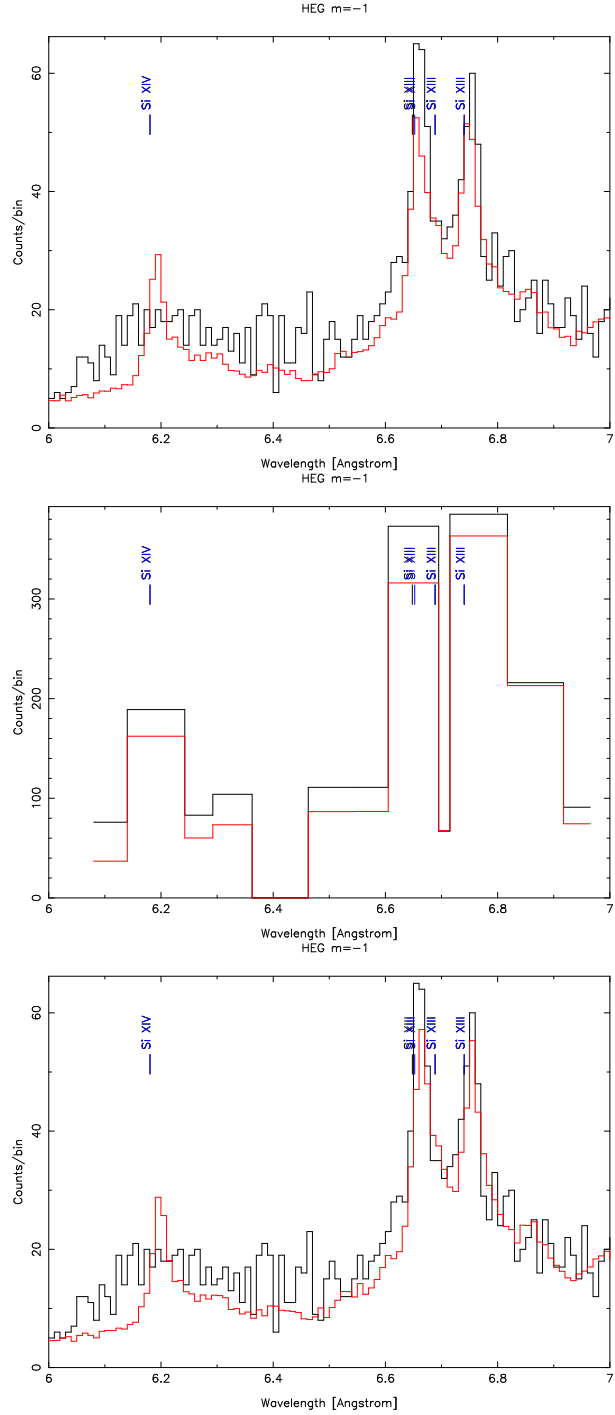

\centering
\includegraphics[height=8cm,angle=-90]{f16a.ps}\\
\includegraphics[height=8cm,angle=-90]{f16b.ps}\\
\includegraphics[height=8cm,angle=-90]{f16c.ps}
\caption{An example of coarse fitting to determine line fluxes.
The top panel shows the data and model fit for the HEG m=$-1$
spectrum using our nominal binning; the HEG m=$+1$ and
MEG m=$\pm 1$ spectra are simultaneously fit as well but
not shown.  This fitting allows the centroids to be accurately measured.
In the middle panel the data and model (with line centroids fixed)
are binned on a coarser custom grid that is aligned with the line locations.
Using this coarse binning the spectra are jointly fit
with a constant-percentage error term since systematic errors
now dominate.
In the bottom panel the coarse-fit model and data are
shown on the nominal grid; note that the modelled lines have grown. }
\label{fig-coarse}
\end{figure}

   The mismatch of the zero-velocity line shape used in our RMFs
is more problematic when measuring the line fluxes.
To reduce these mismatch effects 
we can fix the line locations in the model at the centroid values determined
in our nominal fitting and then rebin the data and model to a coarser grid.
We use binning of order 10 bins, with bin boundaries located in the regions
between the lines.  In this way the total counts in a line region
 are compared between model and data while ignoring line shape
differences.  Because of the large number of counts in each coarse bin,
the errors are now dominated by the approximations of our RMFs and
systematic calibration errors between the 4 spectra we are jointly
fitting, HEG m=$\pm 1$ and MEG m=$\pm 1$.
In place of the usual statistical error, we assign a constant error value to the bins
of each coarse spectrum with a value of 4\% of the maximum counts in a bin
of that spectrum.
This produces fits in which the fluxes
of the lines are better estimated, as we confirm by examining
the fits when data and coarser-fit model are re-plotted to our nominal binning,
 as demonstrated in  Figure~\ref{fig-coarse}.

\section{Converting VNEI Parameters to Physical Units}
\label{sec:vnei_to_phys}

  This appendix summarizes the conversion of model parameter values into
physically meaningful plasma quantities, e.g., the electron density, $n_e$,
and the mass of each element present, $M(Z)$.
A variety of models in the XSPEC library, including the VNEI model we
use in this work, digest the properties of the emitting plasma
into a normalization factor, $X_{\rm norm}$, and a set of relative elemental abundances,
$X_A(Z)$. In addition to these, two other source parameters
are needed: the source distance,

\begin{equation}
  d \ [{\rm cm}] = 3.1\times 10^{21} \ d_{\rm kpc} ,
\end{equation}

\noindent and the volume of the emitting region,

\begin{equation}
   V_R \ [{\rm cm^3}] = V_{\rm as3} \ ( d {{\pi}\over{180}} {{1}\over{3600}} )^3 .
\end{equation}

\noindent For convenience the conversion from values in kpc, $d_{\rm kpc}$,
and arcseconds$^3$, $V_{\rm as3}$, are shown in these equations.

  The actual number fraction of the element $Z$ in the plasma is then given
by:

\begin{equation}
   f(Z) = X_A(Z) \ A_{\rm model}(Z) \ / \ \sum X_A(Z) A_{\rm model}(Z) , 
\end{equation}

\noindent where $A_{\rm model}(Z)$ is the reference ``solar'' number abundance ratio assumed
by the model;
in our case these are the \citet{anders89} values.
The parameters, $X_A(Z)$, are the usual relative abundance values
as input in XSPEC models.

  Because we do not assume that hydrogen dominates the plasma, it is necessary for
the electron density to be self-consistent with the densities and ionization
states of the ions present.  The ratio of electron $n_e$ to total ion density $n_i$ is
then given by:

\begin{equation}
   ({ {n_e} \over {n_i}}) \ = \ \sum Q(Z)\ f(Z) ,
\end{equation}

\noindent where $Q(Z)$ is the average number of electrons stripped from the
ions of element $Z$, in the range 0 to $Z$.  Ideally $Q(Z)$ would be
provided by the model; lacking direct access to it, however, 
we can make a simple  approximation to the ionization state of the elements
in our O-rich plasma by setting:

\begin{equation}
Q(Z) \ = \cases{Z,& if $Z \leq 9$; \cr
              Z-2,& for $10 \leq  Z \leq 16$; \cr
             Z-10,& if $Z \geq 17$. \cr }
\end{equation}

  The electron density can then be calculated as:

\begin{equation}
    n_e = \sqrt{ X_A(Z=1) \ ( n_e n_H )  \ ({ {n_e} \over {n_i}})  \  {{1}\over{f(Z=1)}}     } ,
\end{equation}

\noindent where $n_H$ is the hydrogen density and $( n_e n_H )$ is given by the usual normalization definition

\begin{equation}
     ( n_e n_H ) \ = \ 4 \pi d^2 \  {10^{14}} \   X_{\rm norm}  /  V_R .
\end{equation}

\noindent The $X_A(Z=1)$ factor is included in case the hydrogen
model abundance is set to other than 1.0, e.g., to a small value like 1\ee{-9}
for a pure metal plasma.  The density of element $Z$ is then given by

\begin{equation}
    n(Z) \  = \ f(Z) \  n_e \ / \ ({ {n_e} \over {n_i}}) ,
\end{equation}

\noindent and other quantities like the mass or mass fraction
can be calculated in a straight forward way using the volume, $V$,
and appropriate constants (1~amu = 1.66\ee{-24}~g and \msol = 2\ee{33}~g.)



\clearpage

\begin{deluxetable}{ccccccc}
\tablewidth{0pt}
\tablecaption{Measured and derived parameters for 17 Cas~A regions from HEG+MEG spectral fits to Si lines: line velocities, plasma temperature $kT$, ionization timescale $\tau$, and Si line ratios, forbidden to recombination line ratio and XIV to XIII line ratio. We also list the  type of line profile in the last column: "n" indicates narrow, normal, double peaked Si XIII line
profile produced by $r$ and $f$ lines, and ``sm'' indicates smeared line profile 
due to velocity gradient. Note that for the three regions given in 
parenthesis the  Si ratios are not on the model grid (see Fig.~\ref{fig-line_ratios}) and so the
$kT$ and $\tau$ values assigned are approximate.\label{tab-hetg}} 
\tablehead{ Region & Velocity\tablenotemark{a} & $kT$ & $\tau$ & Si  & Si   & Line \\
  & (\kms)  &  (keV) & (\cms) & (f/r) &  {\small (XIV/XIII)}  & profile }
\startdata
 R1 & $-2600\pm 70$ & 0.77$^{+0.04}_{-0.04}$ & 2.1$^{+2.1}_{-0.8}$\ee{11} & 
$0.59 \pm 0.05$ & $0.07 \pm 0.02$ & n \\ 
 R2 & $-1715\pm 80$  & 1.09$^{+0.24}_{-0.05}$ & 1.1$^{+0.4}_{-0.5}$\ee{11} & 
$0.44 \pm 0.05$ & $0.12 \pm 0.02$ & n \\ 
 R3 & $-380\pm 80$   & 0.74$^{+0.07}_{-0.07}$ & 4.2$^{+22}_{-2.9}$\ee{11} & 
$0.61 \pm 0.06$ & $0.12 \pm 0.03$ & n \\
 R4 & $-620\pm 150$   & 0.74$^{+0.11}_{-0.10}$ & 2.1$^{+5.8}_{-1.3}$\ee{11} & 
$0.60 \pm 0.07$ & $0.06 \pm 0.03$ & n \\
 R5 & $-1735\pm 118$ & 0.77$^{+0.13}_{-0.07}$ & 3.6$^{+5.7}_{-2.1}$\ee{11} & 
$0.59 \pm 0.07$ & $0.12 \pm 0.03$ & sm \\
 (R6) & $-1490\pm 90$  & 0.43 ($< 1.5$) & 3.5\ee{9} ($> 2$\ee{9}) & 
$0.86 \pm 0.10$ & $ 0.001 \pm 0.027$ & sm \\
 R7 & $+3585\pm 135$  & 1.20$^{+0.20}_{-0.16}$ & 4.9$^{+4.3}_{-2.5}$\ee{10} & 
$0.41 \pm 0.04$ & $0.05 \pm 0.03$ & n \\
 R8 & $+2360\pm 140$  & 4.80$^{+1.00}_{-1.1}$ & 2.2$^{+0.6}_{-0.6}$\ee{10} & 
$0.10 \pm 0.04$ & $0.13 \pm 0.04$ & sm \\
 (R9) & $-1150\pm 90$  & $< 1.30$ & $> 6$\ee{11} & 
$0.70 \pm 0.17$ & $0.54 \pm 0.10$ & sm? \\
R10 & $+2700\pm 650$   & 4.30$^{+0.50}_{-0.20}$ & 1.3$^{+0.3}_{-0.5}$\ee{10} & 
$0.11 \pm 0.01$ & $0.05 \pm 0.02$ & sm \\
R11 & $+310\pm 250$      & 0.99$^{+0.06}_{-0.14}$ & 3.6$^{+11}_{-1.8}$\ee{11} & 
$0.54 \pm 0.08$ & $0.29 \pm 0.06$ & n \\
(R12) & $-850\pm 85$   & $>0.81$ & $>2.6$\ee{12} & 
$0.97 \pm 0.13$ & $0.17 \pm 0.04$ & n \\
R13 & $-1070\pm 140$  & 0.90$^{+0.25}_{-0.13}$ & 3.6$^{+11.3}_{-2.3}$\ee{11} & 
$0.54 \pm 0.08$ & $0.20 \pm 0.04$ & n \\
R14 & $+4100\pm 170$  & 0.99$^{+0.21}_{-0.09}$ & 2.4$^{+2.1}_{-1.3}$\ee{11} & 
$0.50 \pm 0.07$ & $0.20 \pm 0.05$ & n \\
R15 & $+760\pm 210$    & 1.00$^{+0.20}_{-0.01}$ & 2.4$^{+0.6}_{-1.4}$\ee{11} & 
$0.47 \pm 0.04$ & $0.17 \pm 0.03$ & sm \\
R16 & $-1420\pm 220$  & 1.46$^{+0.61}_{-0.26}$ & 1.2$^{+1.0}_{-0.6}$\ee{11} & 
$0.37 \pm 0.08$ & $0.28 \pm 0.06$ & n \\
R17 & $-2570\pm 360$  & 4.30$^{+3.70}_{-1.40}$ & 7.3$^{+4.4}_{-2.0}$\ee{10} & 
$0.14 \pm 0.10$ & $0.96 \pm 0.18$ & sm \\
\enddata
\tablenotetext{a}{The velocity
values are obtained using full resolution data, in contrast to the
other results.}
\end{deluxetable}

\clearpage

\input{tab2_rev}

\clearpage

\begin{deluxetable}{ccccc}
\tablewidth{0pt}
\tablecaption{Parameters derived from the zeroth order spectra for the 17 defined regions in Figure~\ref{fig-regions}: electron density $n_e$, total ejecta mass $M_{\rm total}$, time since region was shocked $t_{\rm shock}$, and oxygen mass fraction $M_{\rm oxygen}$/$M_{\rm total}$. As in Table~\ref{tab-hetg}, for the three regions given in 
parenthesis the accuracy of the derived values depends on the model assumed. \label{tab-zo-param}} 
\tablehead{ Region & $n_e$ & $M_{\rm total}$ & $t_{\rm shock}$ & $M_{\rm oxygen}$ \\
   &  (\cm{-3}) & (\ee{-3}\msol) & (yr) & /$M_{\rm total}$}
\startdata
R1  &    93   &  1.3  &    71  &   0.82 \\
R2  &    81   &  0.6  &    43  &   0.90 \\
R3  &   161   &  0.7  &    83  &   0.94 \\
R4  &   157   &  0.7  &    42  &   0.94 \\
R5  &   235   &  0.3  &    48  &   0.93 \\
(R6)  &   203   &  0.3  &   0.5  &   0.95 \\
R7  &    82   &  0.2  &    19  &   0.88 \\
R8  &    76   &  0.1  &     9  &   0.91 \\
(R9)  &    89   &  0.5  &   $>$200  &   0.90 \\
R10 &    32   &  0.4  &    13  &   0.88 \\
R11 &    98   &  0.5  &   116  &   0.95 \\
(R12) &   140   &  0.1  &    $>$600  &   0.92 \\
R13 &   101   &  0.5  &   112  &   0.92 \\
R14 &   131   &  0.3  &    58  &   0.97 \\
R15 &   123   &  0.3  &    62  &   0.96 \\
R16 &    47   &  0.9  &    80  &   0.92 \\
R17 &    23   &  0.3  &   101  &   0.90 \\
\enddata
\end{deluxetable}


\clearpage

\begin{figure}
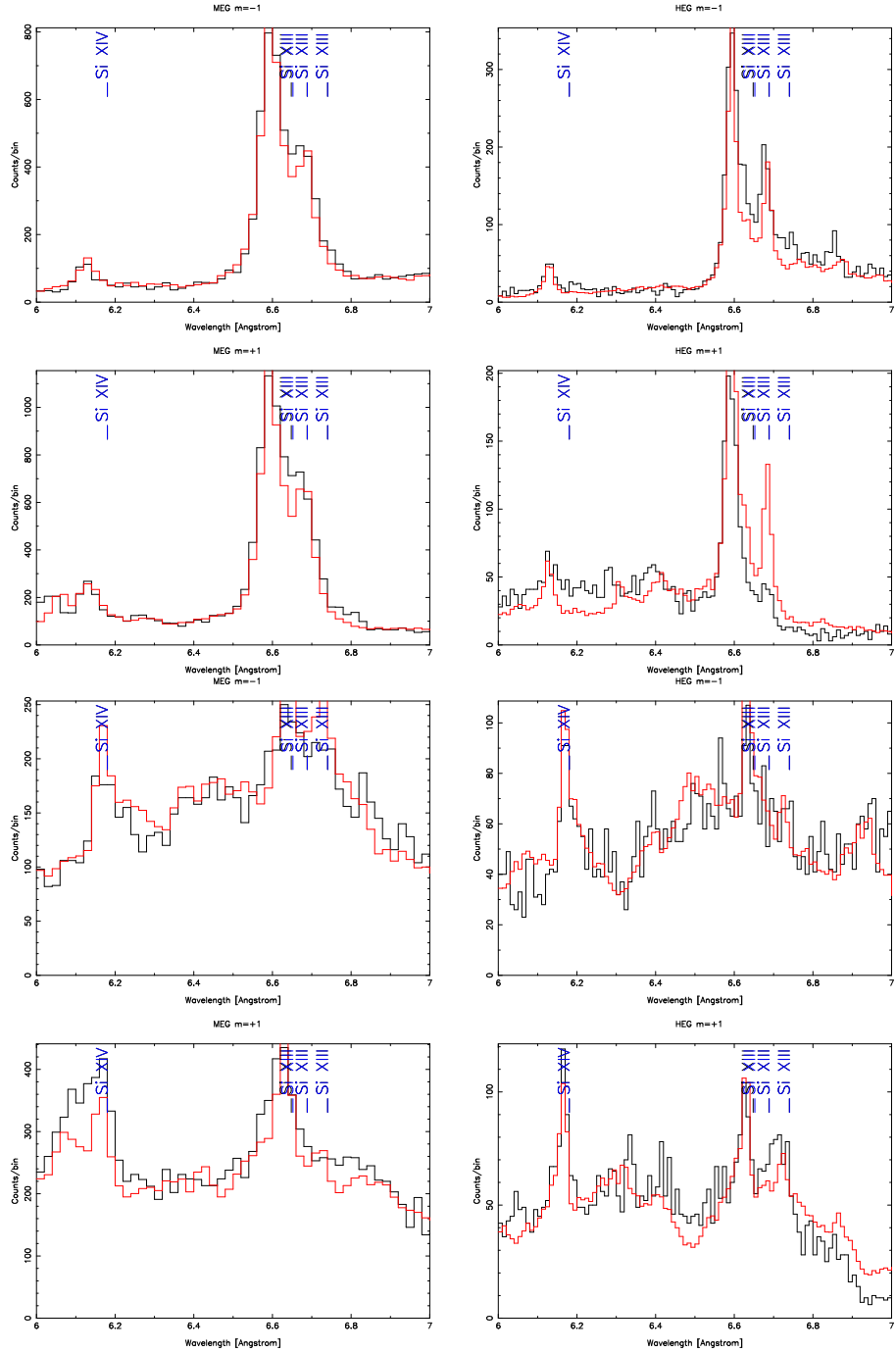

\centering
\includegraphics[height=12cm,angle=-90]{f4a.ps}
\includegraphics[height=12cm,angle=-90]{f4b.ps}
\caption{Examples of HETGS spectra of the Si band encompassing four Si lines used in Cas~A analysis. The upper four panels show the MEG$\pm 1$ (left) and
HEG$\pm 1$ (right) spectra from region R1. The model is plotted with the red line.  The nominal positions of Si lines are marked at the top of each panel. For region R1 the observed lines are shifted to lower wavelength, so this region is an example of a blue-shifted region. The lower four panels similarly show the spectra from region R9; this region has
a very high Si XIV to Si XIII ratio. Region R1 is also an example of a strong, clear line profile produced by a well isolated clump  and little of velocity smearing, whereas from region R9 the line profiles are messy due to lower contrast between the clump and the surrounding medium and possible velocity smearing. }
\label{fig-hetg-si}
\end{figure}

\end{document}

%% file: tab2_rev.tex
\begin{deluxetable}{ccccccc}
\tablewidth{0pt}
\tablecaption{Emission volume for 17 Cas~A regions, $V_R$, and measured parameters from VNEI fits to the zeroth-order spectra: VNEI normalisation factor, $X_{\rm norm}$, and elemental abundances with respect to oxygen (the abundance of O is taken to be
1000 its solar value). Note that zero abundance ratio values denote ratio values smaller than 0.01.  As in Table~\ref{tab-hetg}, for the three regions given in 
parenthesis the accuracy of the derived values depends on the model assumed.\label{tab-zo}} 
\tablehead{ Region & $V_R$ & $X_{\rm norm}$ & Mg/O & Si/O & S/O & Ca/O  \\ 
   & (\ee{50}\cm{3}) & (\ee{-5}\cm{-5}) &  &  &  &    }
\startdata

       R1 &          83 & $6.4_{-0.8}^{+0.5}$ & 0.0 &
$1.5_{-0.1}^{+0.2}$
 &
$2.6_{-0.1}^{+0.2}$
 &
$1.8_{-0.7}^{+1.3}$
 \\
       R2 &          44 & $2.8_{-0.5}^{+0.3}$
 &
$0.0$
 &
$0.8_{-0.1}^{+0.1}$
 &
$1.3_{-0.2}^{+0.2}$
 &
$0.9_{-0.4}^{+0.8}$
 \\
       R3 &          25 & $6.6_{-0.6}^{+0.5}$
 &
$0.04_{-0.03}^{+0.03}$
 &
$0.5_{-0.1}^{+0.1}$
 &
$0.6_{-0.1}^{+0.1}$
 &
$0.7_{-0.5}^{+0.6}$
 \\
       R4 &          28 & $7.0_{-0.6}^{+0.5}$
 &
$0.02_{-0.02}^{+0.02}$
 &
$0.4_{-0.1}^{+0.1}$
 &
$0.6_{-0.1}^{+0.1}$
 &
$0.8_{-0.5}^{+0.6}$
 \\
       R5 &           6 & $3.8_{-0.5}^{+0.1}$
 &
0.0
 &
$0.5_{-0.1}^{+0.1}$
 &
$0.8_{-0.1}^{+0.2}$
 &
$0.4_{-0.3}^{+1.2}$
 \\
       (R6) &          10 & $4.3_{-0.5}^{+0.4}$
 &
0.0
 &
$0.3_{-0.1}^{+0.1}$
 &
$0.6_{-0.1}^{+0.1}$
 &
$0.03_{-0.03}^{+0.04}$
 \\
       R7 &          13 & $0.9_{-0.1}^{+0.1}$
 &
$0.1_{-0.1}^{+0.1}$
 &
$0.8_{-0.1}^{+0.1}$
 &
$1.5_{-0.2}^{+0.1}$
 &
$2.2_{-0.9}^{+0.8}$
 \\
       R8 &           7 & $0.4_{-0.1}^{+0.1}$
 &
$0.2_{-0.1}^{+0.1}$
 &
$0.8_{-0.2}^{+0.3}$
 &
$0.7_{-0.2}^{+0.3}$
 &
$0.4_{-0.4}^{+0.8}$
 \\
       (R9) &          31 & $2.4_{-0.4}^{+0.3}$
 &
$0.3_{-0.1}^{+0.1}$
 &
$0.7_{-0.1}^{+0.1}$
 &
$0.9_{-0.1}^{+0.2}$
 &
$0.5_{-0.3}^{+0.3}$
 \\
      R10 &          65 & $0.7_{-0.2}^{+0.1}$
 &
$0.2_{-0.1}^{+0.1}$
 &
$1.0_{-0.2}^{+0.3}$
 &
$1.2_{-0.3}^{+0.4}$
 &
$0.1_{-0.1}^{+0.3}$
 \\
      R11 &          29 & $2.9_{-0.3}^{+0.4}$
 &
$0.1_{-0.3}^{+0.4}$
 &
$0.3_{-0.1}^{+0.1}$
 &
$0.5_{-0.1}^{+0.1}$
 &
$0.1_{-0.1}^{+0.4}$
 \\
      (R12) &           5 & $1.1_{-0.6}^{+0.2}$
 &
0.0
 &
$0.9_{-0.1}^{+0.6}$
 &
$2.0_{-0.8}^{+1.6}$
 &
0.0
 \\
      R13 &          27 & $2.8_{-0.5}^{+0.5}$
 &
$0.05_{-0.05}^{+0.06}$
 &
$0.5_{-0.1}^{+0.1}$
 &
$1.1_{-0.2}^{+0.3}$
 &
$1.1_{-0.7}^{+0.8}$
 \\
      R14 &          14 & $2.5_{-0.3}^{+0.4}$
 &
$0.02_{-0.02}^{+0.04}$
 &
$0.2_{-0.1}^{+0.1}$
 &
$0.3_{-0.1}^{+0.1}$
 &
$0.6_{-0.5}^{+0.6}$
 \\
      R15 &          16 & $2.5_{-0.4}^{+0.1}$
 &
0.0
 &
$0.4_{-0.1}^{+0.1}$
 &
$0.5_{-0.1}^{+0.2}$
 &
0.0
 \\
      R16 &         109 & $2.4_{-0.4}^{+0.3}$
 &
$0.4_{-0.1}^{+0.1}$
 &
$0.6_{-0.1}^{+0.1}$
 &
$0.5_{-0.1}^{+0.1}$
 &
$0.2_{-0.2}^{+0.3}$
 \\
      R17 &          87 & $0.4_{-0.1}^{+0.2}$
 &
$0.7_{-0.2}^{+0.3}$
 &
$0.6_{-0.1}^{+0.3}$
 &
$0.5_{-0.1}^{+0.2}$
 &
$0.2_{-0.2}^{+0.6}$
 \\
\enddata
\end{deluxetable}